\documentclass[english,aps,manuscript]{revtex4}
\usepackage[T1]{fontenc}
\usepackage[latin9]{inputenc}
\usepackage[letterpaper]{geometry}
\usepackage{amsmath}
\usepackage{graphicx}
\usepackage{amssymb}
\usepackage{esint}

\makeatletter

\newcommand{\noun}[1]{\textsc{#1}}
\providecommand{\tabularnewline}{\\}

\@ifundefined{textcolor}{}
{%
 \definecolor{BLACK}{gray}{0}
 \definecolor{WHITE}{gray}{1}
 \definecolor{RED}{rgb}{1,0,0}
 \definecolor{GREEN}{rgb}{0,1,0}
 \definecolor{BLUE}{rgb}{0,0,1}
 \definecolor{CYAN}{cmyk}{1,0,0,0}
 \definecolor{MAGENTA}{cmyk}{0,1,0,0}
 \definecolor{YELLOW}{cmyk}{0,0,1,0}
 }

\makeatother

\usepackage{babel}

\begin{document}

\title{Quantum process tomography of molecular dimers from two-dimensional
electronic spectroscopy I: General theory and application to homodimers}

\author{Joel Yuen-Zhou and Alán Aspuru-Guzik}

\email{aspuru@chemistry.harvard.edu}

\address{Department of Chemistry and Chemical Biology, Harvard University,
Cambridge, MA 02138 }
\begin{abstract}
Is it possible to infer the time evolving quantum state of a multichromophoric
system from a sequence of two-dimensional electronic spectra (2D-ES)
as a function of waiting time? Here we provide a positive answer for
a tractable model system: a coupled dimer. After exhaustively enumerating
the Liouville pathways associated to each peak in the 2D-ES, we argue
that by judiciously combining the information from a series of experiments
varying the polarization and frequency components of the pulses, detailed
information at the amplitude level about the input and output quantum
states at the waiting time can be obtained. This possibility yields
a quantum process tomography (QPT) of the single-exciton manifold,
which completely characterizes the open quantum system dynamics through
the reconstruction of the process matrix. This is the first of a series
of two articles. In this manuscript, we specialize our results to
the case of a homodimer, where we prove that signals stemming from
coherence to population transfer and viceversa vanish upon isotropic
averaging, and therefore, only a partial QPT is possible in this case.
However, this fact simplifies the spectra, and it follows that only
two polarization controlled experiments (and no pulse-shaping requirements)
suffice to yield the elements of the process matrix which survive
under isotropic averaging. The angle between the two site transition
dipole moments is self-consistently obtained from the 2D-ES. Model
calculations are presented, as well as an error analysis in terms
of the angle between the dipoles and peak overlap. In the second article
accompanying this study, we numerically exemplify the theory for heterodimers
and carry out a detailed error analysis for such case. This investigation
provides an important benchmark for more complex proposals of quantum
process tomography (QPT) via multidimensional spectroscopic experiments.
\end{abstract}
\maketitle
Multidimensional optical spectroscopies (MDOS) provide very powerful
tools to study excited state dynamics of multichromophoric systems
in condensed phases. These techniques distribute spectral features
along several dimensions, uncluttering data which would otherwise
appear obscured in linear spectroscopies and simultaneously yielding
novel information on the dynamics of the probed system \cite{mukamel}.
Possibilities in multidimensional techniques include decongesting
spectral lineshapes, differentiating between homogeneous and inhomogeneous
broadening mechanisms, providing unambiguous signatures about couplings
between chromophores, and yielding signatures of coherent and incoherent
processes involving excited states at the amplitude level \cite{minhaengbook,minhaeng}.
Although MDOS have been historically inspired by their NMR analogues,
the timescales of the physical and chemical processes studied through
MDOS are quite different from the ones in NMR \cite{mukamelacc,jonas,mukamel_accchemres,harel}.
The characteristic timescales of NMR are milliseconds, a resolution
that does not allow for the observation of a wide variety of chemical
dynamics in condensed phases ocurring in the orders of femto and picoseconds.
On the other hand, femtosecond timescales can be easily accessed with
ultrafast optical techniques. Examples of phenomena studied via MDOS
are vast and include molecular reorientation processes and solvation
dynamics \cite{prlliquids,fayer-accchemres}, electron transfer \cite{moran-et},
vibrational coherences in organometallic complexes \cite{khalil:362,khaliljpc,Baiz2009}
or halogens in rare gas matrices \cite{bihary,guhr-bromine}, phonon
dynamics in carbon nanotubes \cite{nanotube-uosaki}, protein unfolding
kinetics \cite{tokmakoffprotein}, excitonic dynamics in light-harvesting
systems \cite{engelchicago,engelfleming,scholes,moran,hauer_carotene,kauffmann_accchemres}
and organic polymers \cite{ppv,collinischoles}, as well as many-body
physics in semiconductor quantum wells \cite{Stone05292009,turner-nelson,turnerjcp}
and quantum dots \cite{Wong2010}.

Traditionally, the spectroscopy of condensed phases is formulated
as a response problem: The molecular system is perturbed with a sequence
of short laser pulses and the coherent polarization response due to
this set of perturbations (nonlinear polarization) is subsequently
measured \cite{mukamel}. Informally, we can describe the exercise
as 'kicking' the quantum black box (molecular system) and 'listening
to the whispers' (measuring the response) due to the kicks, from which
some properties of the box can be inferred. This description of spectroscopy
is reminiscent to an idea stemming from the quantum optics and quantum
information processing (QIP) communities, namely, quantum process
tomography (QPT) \cite{nielsenchuangbook,nielsenchuang,dcqd,resources}.
Broadly speaking, QPT is a systematic procedure to characterize a
quantum black box by sending a set of inputs, measuring their outputs,
and analyzing the functional relationships between them. With the
increasing effort of quantum engineering of gates and devices, QPT
constitutes a cornerstone of QIP theory and experiment, as it provides
a necessary check on the performance of the respective quantum black
boxes. A natural question arises from the comparison of the two aforementioned
concepts: Can the spectroscopy of condensed phases be formulated as
a QPT? In a previous study \cite{yuenzhou}, we provided an affirmative
answer to this question, at least for a molecular dimer. We showed
that a set of two-color polarization controlled rephasing photon echo
experiments is sufficient to reconstruct the density matrix elements
associated with the dynamics of the single exciton manifold, and therefore,
systematically characterize the excited state dynamics of the dimer,
which can be regarded as the black box. For pedagogical reasons, we
found it simpler and more convenient to concentrate our attention
in the real time picture of the experiment, to make an explicit identification
of the preparation, evolution, and detection steps of the QPT, with
the coherence, waiting, and echo times, respectively. However, due
to the widespread practice of displaying partially Fourier transformed
data of the nonlinear optical polarization with respect to certain
time intervals, it is worthwhile translating our results to the more
visual two-dimensional electronic spectrum (2D-ES), and in fact, this
is one of the main results of this work.

The present article is organized as follows: We begin in section I
with a review of some relevant ideas of QPT and also introduce the
process matrix as the main object to be reconstructed by means of
QPT. In section II, relevant details on the dimer model system are
presented. Section III describes the rephasing heterodyne photon echo
experiment for the dimer and explains that the collected macroscopic
polarization signal is a linear combination of elements of the process
matrix at the waiting time $\chi(T)$. This implies that QPT can be
performed by repeating several experiments with different pulse parameters
in order to extract these elements. In section IV, the ideas of section
III are mapped into the language of a 2D-ES, where each of the diagonal
and cross peaks is associated with a set of elements of $\chi(T)$,
and each of the axes of the spectrum can be associated with a preparation
and a detection stage. Finally, in section V, these ideas are specialized
to homodimer systems where, after isotropic averaging, only a partial
QPT is possible, as some elements of $\chi(T)$ are undetectable.
Nevertheless, we note that the partial QPT is easily realized with
current experimental capabilities, since it can be reconstructed with
only two spectra resulting from different pulse polarization configurations
for each given waiting time. The angle between the site dipoles is
self-consistently obtained from these spectra, and an error analysis
based on this angle as well as peak overlaps is carried out. Numerical
calculations on a secular Redfield dissipation model are presented.
A detailed analysis for heterodimers is carried out in the next article
accompanying this investigation. Extensions of the procedure to account
for inhomogeneous broadening, more sophisticated signal analysis,
as well as bigger systems, are discuss at the end of this manuscript.

\section{Relevant concepts of Quantum Process Tomography and general definitions}

Consider an arbitrary quantum system (quantum black box) interacting
with an environment. We are interested in its evolution as a function
of time $T$ in the form of a reduced density matrix $\rho(T)$. Very
generally, this evolution is a linear transformation on the initial
quantum state \cite{Choi1975285,sudarshan}: 

\begin{equation}
\rho(T)=\chi(T)\rho(0).\label{eq:linear transformation}\end{equation}
$\chi(T)$ is the central object of this article, and shall be called
\emph{process matrix}. Eq. (\ref{eq:linear transformation}) can be
regarded as an integrated equation of motion for every $T$ %
\footnote{We label time with $T$ instead of $t$ because the QPT we propose
is identified with the waiting time $T$ of the PE experiment.%
}. Eq. (\ref{eq:linear transformation}) can be expressed in terms
of a basis for the Liouville space of the system:

\begin{equation}
\rho_{ab}(T)=\sum_{cd}\chi_{abcd}(T)\rho_{cd}(0).\label{eq:linear transformation basis}\end{equation}

For purposes of this article, we present two useful definitions. Consider
the Liouville space $\mathcal{L}$ of the system, and classify the
vectors of $\mathcal{L}$ in proper and improper density matrices.
A state or a density matrix is \emph{proper }if it satisfies all the
conditions of a physical quantum state; namely, this is Hermitian,
positive semidefinite, and has trace one. An \emph{improper} state
is any matrix that lives in the Liouville space but is not a proper
density matrix. Clearly, any improper density matrix in the same Liouville
space may be written as a unique linear combination of proper density
matrices. In principle, Eq. (\ref{eq:linear transformation basis}),
being a physical equation of motion, is restricted to the domain of
proper density matrices $\rho(0)$. However, by linearity, its extension
to any linear combination of proper states is well defined, so its
validity for improper density matrices is not under question. 

The meaning of the process matrix $\chi(T)$ is easy to grasp: Conditional
on the initially state being prepared at $\rho(0)=|c\rangle\langle d|$,
$\chi_{abcd}(T)$ is the value of the entry $ab$ of the quantum state
after time $T$, $\rho(T)$, i.e. $\chi_{abcd}(T)=\langle a|\rho(T)|b\rangle$.
Therefore, $\chi_{abcd}(T)$ denotes a state to state transfer amplitude.
Note that $\rho(0)=|c\rangle\langle d|$ is an improper density matrix
if $c\neq d$ (coherences on their own are not valid quantum states).
However, improper states are not necessarily unphysical as one expects
at a first glance. Most of our intuition for nonlinear spectroscopies
in the perturbative regime stems from the consideration of how a perturbative
amplitude created at a certain entry $|c\rangle\langle d|$ of the
total (proper) density matrix is transferred to other entries due
to free evolution, as time progresses \cite{mukamel}. It is not the
evolution of the total density matrix (which to leading order is unperturbed,
mostly in its ground state, and not yielding a time-dependent dipole)
what is effectively monitored in the phase-matched signal, but the
evolution of an \emph{effective} density matrix, such as $|c\rangle\langle d|$,
which can be improper. Terms such as transfer from population to population,
coherence to coherence, population to coherence, and coherence to
population are all ubiquitous in the jargon of MDOS. However, the
monitoring of the latter is often ambiguous, incomplete, and in most
cases, qualitative. Obtaining quantitative information about these
events amounts to finding each of the elements of $\chi(T)$.

The transformation in Eq. (\ref{eq:linear transformation basis})
is limited by two classes of restrictions for the process matrix associated
with Hermiticity and trace preservation:

\begin{eqnarray}
\sum_{a}\chi_{aacd}(T) & = & \delta_{cd},\label{eq:trace}\\
\chi_{abcd}(T) & = & \chi_{badc}^{*}(T).\label{eq:hermiticity-1}\end{eqnarray}
We derive these conditions in Appendix A, but their content is intuitive:
If $\rho(0)$ is a proper density matrix, $\rho(t)$ remains as a
valid quantum state as $T$ evolves if these two requirements are
preserved. In particular, elements of the form $\chi_{aabb}(T)$,
which denote population transfers, are purely real as one expects,
whereas the other elements are in general complex. As a comment to
our previous discussion, by linearity, these conditions must also
be satisfied even if $\rho(0)$ is improper (notice that $\chi(T)$
does not depend on $\rho(0)$).

Equations (\ref{eq:linear transformation}) and (\ref{eq:linear transformation basis})
are remarkable because they guarantee that, in principle, if $\chi(T)$
is known, the quantum black box described by $\rho(T)$ is perfectly
understood, as it predicts by linearity the evolution of an arbitrary
initial state of $\mathcal{L}$. Although $\rho(t)$ describes an
open quantum system, details about the environment evolution need
not be included explicitly, but only in an averaged sense in the elements
of $\chi(T)$. We shall operationally\emph{ define QPT as any procedure
to reconstruct $\chi(T)$}. A possible QPT is the following: (a) Prepare
a linearly independent set of states $\rho(0)$ that spans $\mathcal{L}$;
(b) for each of the prepared states, wait for a free evolution time
$T$ and determine the density matrix at that time. Any protocol for
determining a density matrix for a system is called Quantum State
Tomography (QST) \cite{statetomography,walmsley,holography,cinaprl,cinapolyatomic}.
In essence, QPT can be carried out for any system if both a selective
preparation of initial states and QST can be achieved. Variants of
this methodology exist although all of them operate within the same
spirit \cite{nielsenchuangbook,nielsenchuang,dcqd,resources}. QPT
has been successfully implemented in a wide variety of experimental
scenarios, including nuclear magnetic resonance \cite{cory,childs,veeman},
ion traps \cite{blatttraps}, single photons \cite{whitephotons,steinbergphotons},
solid state qubits \cite{solidstate}, optical lattices \cite{opticallattices},
and Josephson junctions \cite{natphysqpt}. In this article, we show
how to perform QPT for a model coupled heterodimer using two-color
polarization controlled heterodyne photon-echo experiments, extending
the domain of application of QPT to systems of chemical and biophysical
interest.

\section{Model system: Coupled dimer}

Consider a molecular dimer described by the effective Hamiltonian
\cite{minhaeng,yang:2983,pullerits-jcp}:

\begin{equation}
H_{S}=\omega_{A}a_{A}^{+}a_{A}+\omega_{B}a_{B}^{+}a_{B}+J(a_{A}^{+}a_{B}+a_{B}^{+}a_{A}),\label{eq:excitonic H not diagonalized-1}\end{equation}
where $a_{i}^{+}$ and $a_{i}$ are creation and anhilation operators
for a single Frenkel exciton in the site $i\in\{A,B\}$, $\omega_{A},\omega_{B}$
are the first and second site energies, and $J$ is the coupling between
the chromophores. 

The standard diagonalization of this Hamiltonian, which is effectively
a two-level system for the single-exciton manifold, follows from defining
some convenient parameters: The average of the site energies $\bar{\omega}=\frac{1}{2}(\omega_{A}+\omega{}_{B})$,
the difference $\Delta=\frac{1}{2}(\omega_{A}-\omega_{B})$, and the
mixing angle $\theta=\frac{1}{2}\arctan\left(\frac{J}{\Delta}\right)$.
By introducing the operators:

\begin{align}
a_{\alpha} & =\cos\theta a_{A}+\sin\theta a_{B},\nonumber \\
a_{\beta} & =-\sin\theta a_{A}+\cos\theta a_{B},\label{eq:transformation}\end{align}
the Hamiltonian in Equation (\ref{eq:excitonic H not diagonalized-1})
can be readily written as:

\begin{equation}
H_{S}=\omega_{\alpha}a_{\alpha}^{+}a_{\alpha}+\omega_{\beta}a_{\beta}^{+}a_{\beta},\label{eq:excitonic H-1}\end{equation}
where the eigenvalues $\omega_{\alpha}$ and $\omega_{\beta}$ of
the single excitons are:

\begin{align}
\omega_{\alpha} & =\omega+\Delta\sec2\theta,\nonumber \\
\omega_{\beta} & =\omega-\Delta\sec2\theta.\label{eq:energies}\end{align}
Denoting $|g\rangle$ as the molecular ground state or the excitonic
vaccuum, $|A\rangle=a_{A}^{+}|g\rangle$ and $|B\rangle=a_{\beta}^{+}|g\rangle$
are the excitons at each site, whereas $|\alpha\rangle=a_{\alpha}^{+}|g\rangle$,
$|\beta\rangle=a_{\beta}^{+}|g\rangle$ are the delocalized excitons.
The biexcitonic state, expressed by $|f\rangle=a_{A}^{+}a_{B}^{+}|g\rangle=c_{\alpha}^{+}c_{\beta}^{+}|g\rangle$,
also plays a role in our study, as it is resonantly accesed through
excited state absorption (ESA) after several pulses. Notice that the
Hamiltonian $H_{S}$ does not contain two-body operators, and therefore
does not account for exciton-exciton binding or repulsion terms, so
the energy level of the biexciton is just the sum of the two exciton
energies, $\omega_{f}=\omega_{\alpha}+\omega_{\beta}=\omega_{A}+\omega_{B}$
\cite{minhaengbook}. Defining $\omega_{ij}\equiv\omega_{i}-\omega_{j}$,
the following relations hold:

\begin{eqnarray}
\omega_{\alpha g} & = & \omega_{f\beta}\nonumber \\
\omega_{\beta g} & = & \omega_{f\alpha}.\label{eq:identities-1}\end{eqnarray}

Since we are concerned with the interaction of the chromophores with
electromagnetic radiation, we make some remarks on the geometry of
the transition dipoles (see Fig. 1). Let $\boldsymbol{\mu}_{ij}=\langle i|\hat{\boldsymbol{\mu}}|j\rangle$.
Assume that the transition dipole moments from the ground to the single
excitons in the site basis are $\mbox{\ensuremath{\boldsymbol{\mu}}}_{gA}=\boldsymbol{\mu}_{Ag}=\boldsymbol{d}_{A}$
and $\mbox{\ensuremath{\boldsymbol{\mu}}}_{gB}=\boldsymbol{\mu}_{Bg}=\boldsymbol{d}_{B}$,
respectively. It follows that the dipole moments $\boldsymbol{\mu}_{ij}$
for $i,j\in\{\alpha,\beta,f\}$ are located in the same plane, but
in general have different magnitudes and directions: 

\begin{eqnarray}
\left[\begin{array}{c}
\mathbb{\boldsymbol{\mu}}_{\alpha g}\\
\boldsymbol{\mu}_{\beta g}\end{array}\right] & = & \left[\begin{array}{cc}
\cos\theta & \sin\theta\\
-\sin\theta & \cos\theta\end{array}\right]\left[\begin{array}{c}
\boldsymbol{d}_{A}\\
\boldsymbol{d}_{B}\end{array}\right]\nonumber \\
\left[\begin{array}{c}
\mathbb{\boldsymbol{\mu}}_{f\alpha}\\
\boldsymbol{\mu}_{f\beta}\end{array}\right] & = & \left[\begin{array}{cc}
\sin\theta & \cos\theta\\
\cos\theta & -\sin\theta\end{array}\right]\left[\begin{array}{c}
\boldsymbol{d}_{A}\\
\boldsymbol{d}_{B}\end{array}\right].\label{eq:dipoles}\end{eqnarray}
In this model, we shall consider $\boldsymbol{\mu}_{ij}=\boldsymbol{\mu}_{ji}$.
As enumerated in our model, dipole mediated transitions only couple
the ground state to the single excitons, and the single excitons to
the biexciton.

\section{Photon-echo experiment as Quantum Process Tomography}

Consider a four-wave mixing experiment where an ensemble of identical
dimers interacts with a series of three ultrashort laser pulses. The
perturbation due to these pulses is given by:

\begin{equation}
V(\boldsymbol{r},t)=-\lambda\sum_{i=1}^{3}\hat{\boldsymbol{\mu}}\cdot\boldsymbol{e_{i}}E(t-t_{i})e{}^{i\boldsymbol{k}_{i}\cdot\boldsymbol{r}-i\omega_{i}(t-t_{i})}+c.c.,\label{eq:perturbation}\end{equation}
where $\lambda$ is the intensity of the electric field, which is
assumed to be weak, $\hat{\boldsymbol{\mu}}$ is the dipole operator,
$\boldsymbol{e_{i}},t_{i},\boldsymbol{k}_{i},\omega_{i}$ denote the
polarization %
\footnote{Hereafter, we use the word \emph{polarization} in two different ways:
To denote (a) the orientation of oscillations of the electric field
and (b) the density of electric dipole moments in a material. The
meaning should be clear by the context. %
}, time center, wavector, and carrier frequency of the $i-th$ pulse,
and $\boldsymbol{r}$ is the position of the center of mass of the
molecule. $E(t)$ is the slowly varying in time pulse envelope, which
we choose as a Gaussian with width $\sigma$, or full-width half-maximum
$FWHM=2\sqrt{2\ln2}\sigma$, $E(t)=e^{-t^{2}/(2\sigma^{2})}$. The
pulses are sent to the sample in a non-collinear fashion to the sample,
generating a time-dependent dipole in each of the molecules. Since
the characteristic size of a molecule is much smaller than the wavelength
of the radiation, $2\pi/|\boldsymbol{k}_{i}|$, each molecule only
experiences a potential that changes in time but is uniform in space,
in consistency with the dipole approximation. Nonetheless, the spatial
dependence of the pulses is still important, as the phases $e{}^{\pm i\boldsymbol{k}_{i}\cdot\boldsymbol{r}}$
are imprinted to molecules located across different positions $\boldsymbol{r}$
in the sample. The size of the sample is much larger than $2\pi/|\boldsymbol{k}_{i}|$,
so there is a considerable spacial modulation of the polarization
due to these phases. Denoting the time-dependent state of the molecule
at position $\boldsymbol{r}$ by $\rho(\boldsymbol{r},t)$, a perturbative
treatment allows us to decompose the density matrix into Fourier components:

\begin{equation}
\rho(\boldsymbol{r},t)=\sum_{s}\rho_{s}(t)e^{i\boldsymbol{k}_{s}\cdot\boldsymbol{r}}\label{eq:fourier density matrix}\end{equation}
where $\boldsymbol{k}_{s}=l\boldsymbol{k}_{1}+m\boldsymbol{k}_{2}+n\boldsymbol{k}_{3}$
and $l,n,m$ are integer numbers. Notice that $\boldsymbol{k}_{s}$
equals to a linear combination of the wavevectors associated with
each pulse. As expected, the action of a pulse on each molecule attaches
a spatial phase to its quantum state, so the total phase accumulated
by it equals $e^{i\boldsymbol{k}_{s}\cdot\boldsymbol{r}}$ for each
combination of perturbations. Each improper density matrix $\rho_{s}(t)$
corresponds to one of these phases, and can be calculated by keeping
track of the actions of the pulses in the bra and the ket of the system
using double-sided Feynman diagrams. Eq. (\ref{eq:fourier density matrix})
implies that the optical polarization induced on the molecule can
also be Fourier decomposed into different components \cite{seidner:3998,pullerits-prb,gelin:164112}:
$\boldsymbol{P}(\boldsymbol{r},t)=Tr(\hat{\boldsymbol{\mu}}(\boldsymbol{r})\rho(\boldsymbol{r},t))=\sum_{s}\boldsymbol{P}_{s}(t)e^{i\boldsymbol{k}_{s}\cdot\boldsymbol{r}}$,
where $\hat{\boldsymbol{\mu}}(\boldsymbol{r})$ denotes the dipole
operator of the molecule located at $\boldsymbol{r}$. The experimental
setting we describe is analogous to the one of an array of dipole
antennas which are spatially phased in a grating with respect to each
other and oscillate in time. Classical electromagnetism predicts that
the induced macroscopic polarization of this array emits radiation
which is precisely concentrated along the vectors $\boldsymbol{k}_{s}$.
This condition, which reflects conservation of momentum of the fields,
is known as\emph{ phase-matching} \cite{echoes}. A fourth pulse of
the same wavevector as one of the $\boldsymbol{k}_{s}$, known as
the \emph{local oscillator}, is allowed to interfere with the radiation
along that direction. By varying the phases of this fourth field,
two heterodyne detections can be carried out to extract the real and
imaginary components of $P_{s}(t)\equiv\boldsymbol{P}_{s}(t)\cdot\boldsymbol{e}_{4}$
respectively, where $\boldsymbol{e}_{4}$ is the polarization of the
local oscillator \cite{mukamel}. 

In this article, we are interested in the signal along $\boldsymbol{k}_{PE}=-\boldsymbol{k}_{1}+\boldsymbol{k}_{2}+\boldsymbol{k_{3}}$,
the so called photon-echo (PE) direction \cite{echoes-prl}. The frequency
components of the pulses lie within the optical regime, so they can
induce the transitions enumerated in the previous section. Traditionally,
in the MDOS literature, the intervals between the time centers of
the pulses are called \emph{coherence} $\tau=t_{2}-t_{1}$, \emph{waiting}
$T=t_{3}-t_{2}$, and \emph{echo} $t=t_{4}-t_{3}$ times, respectively.
Here, $t_{4}$ is the time of detection of the signal \cite{chinesejournal}.
We shall only consider rephasing photon-echo signals, where $t_{1}<t_{2}<t_{3}<t_{4}$,
where the inhomogeneous broadening is rephased \cite{cheng}. Due
to these explicit interval dependences, the collected signal can be
expressed as $P_{PE}(\mbox{\ensuremath{\tau}},T,t)$. As explained
in our previous study \cite{yuenzhou}, we may regard the PE experiment
as a QPT of the single-exciton manifold dynamics of the dimer as a
function of $T$. In fact, the polarization signal may be expressed
as a linear combination of elements of the process matrix $\chi(T)$:

\begin{eqnarray}
[P_{PE}]{}_{\boldsymbol{e}_{1},\boldsymbol{e}_{2},\boldsymbol{e}_{3},\boldsymbol{e}_{4}}^{\omega_{1},\omega_{2},\omega_{3}}(t) & = & \sum_{p,q,r}C_{\omega_{1}}^{p}C_{\omega_{2}}^{q}C_{\omega_{3}}^{r}P{}_{\boldsymbol{e}_{1},\boldsymbol{e}_{2},\boldsymbol{e}_{3},\boldsymbol{e}_{4}}^{p,q,r}(t),\label{eq:total polarization}\end{eqnarray}
 where,

\begin{eqnarray}
 &  & P{}_{\boldsymbol{e}_{1},\boldsymbol{e}_{2},\boldsymbol{e}_{3},\boldsymbol{e}_{4}}^{p,q,\alpha}(t)\nonumber \\
 & = & -(\boldsymbol{\mu}_{pg}\cdot\boldsymbol{e}_{1})(\boldsymbol{\mu}_{qg}\cdot\boldsymbol{e}_{2})\mathcal{G}_{gp}(\tau)\nonumber \\
 &  & \times\{[(\boldsymbol{\mu}_{\alpha g}\cdot\boldsymbol{e}_{3})(\boldsymbol{\mu}_{\alpha g}\cdot\boldsymbol{e}_{4})\mathcal{G}_{\alpha g}(t)\nonumber \\
 &  & \times(\chi_{ggqp}(T)-\delta_{pq}-\chi_{\alpha\alpha qp}(T))\nonumber \\
 &  & +(\boldsymbol{\mu}_{f\beta}\cdot\boldsymbol{e}_{3})(\boldsymbol{\mu}_{f\beta}\cdot\boldsymbol{e}_{4})\mathcal{G}_{f\beta}(t)\chi_{\beta\beta qp}(T)\nonumber \\
 &  & +((\boldsymbol{\mu}_{f\beta}\cdot\boldsymbol{e}_{3})(\boldsymbol{\mu}_{f\alpha}\cdot\boldsymbol{e}_{4})\mathcal{G}_{f\alpha}(t)\nonumber \\
 &  & -(\boldsymbol{\mu}_{\alpha g}\cdot\boldsymbol{e}_{3})(\boldsymbol{\mu}_{\beta g}\cdot\boldsymbol{e}_{4})\mathcal{G}_{\beta g}(t))\chi_{\beta\alpha qp}(T)]\},\label{eq:with alpha}\end{eqnarray}
 and,

\begin{eqnarray}
 &  & P{}_{\boldsymbol{e}_{1},\boldsymbol{e}_{2},\boldsymbol{e}_{3},\boldsymbol{e}_{4}}^{p,q,\beta}(t)\nonumber \\
 & = & -(\boldsymbol{\mu}_{pg}\cdot\boldsymbol{e}_{1})(\boldsymbol{\mu}_{qg}\cdot\boldsymbol{e}_{2})\mathcal{G}_{gp}(t)\nonumber \\
 &  & \times\{[(\boldsymbol{\mu}_{\beta g}\cdot\boldsymbol{e}_{3})(\boldsymbol{\mu}_{\beta g}\cdot\boldsymbol{e}_{4})\mathcal{G}_{\beta g}(t)\nonumber \\
 &  & \times(\chi_{ggqp}(T)-\delta_{pq}-\chi_{\beta\beta qp}(T))\nonumber \\
 &  & +(\boldsymbol{\mu}_{f\alpha}\cdot\boldsymbol{e}_{3})(\boldsymbol{\mu}_{f\alpha}\cdot\boldsymbol{e}_{4})\mathcal{G}_{f\alpha}(t)\chi_{\alpha\alpha qp}(T)\nonumber \\
 &  & +((\boldsymbol{\mu}_{f\alpha}\cdot\boldsymbol{e}_{3})(\boldsymbol{\mu}_{f\beta}\cdot\boldsymbol{e}_{4})\mathcal{G}_{f\beta}(t)\nonumber \\
 &  & -(\boldsymbol{\mu}_{\beta g}\cdot\boldsymbol{e}_{3})(\boldsymbol{\mu}_{\alpha g}\cdot\boldsymbol{e}_{4})\mathcal{G}_{\alpha g}(t))\chi_{\alpha\beta qp}(T)]\}.\label{eq:with beta}\end{eqnarray}

The coefficients $C_{\omega_{i}}^{p}$ for $p\in\{\alpha,\beta\}$
are frequency amplitudes of the laser pulse which is centered at $\omega_{i}$,
evaluated at the transition energy $\omega_{pg}$:

\begin{equation}
C_{\omega_{i}}^{p}=-\frac{\lambda}{i}\sqrt{2\pi\sigma^{2}}e^{-\sigma^{2}(\omega_{pg}-\omega_{i})^{2}/2},\label{eq:coefficient}\end{equation}
 and

\begin{equation}
\mathcal{G}_{ij}(\tau)=\Theta(\tau)e^{(-i\omega_{ij}-\Gamma_{ij})\tau}\label{eq:optical coherence}\end{equation}
is the propagator of the optical coherences $|i\rangle\langle j|$
in the coherence and echo times, which, has been taken to be the product
of a coherent oscillatory term beating at a frequency $\omega_{ij}$
and an exponential decay with dephasing rate $\Gamma_{ij}$. This
propagator is defined only for $\tau>0$ via the step function $\Theta(\tau)$.
The frequencies of the coherences in the coherence and echo intervals
have opposite signs, reflecting the rephasing character of the signal.
In optical PE experiments, it is customary to assume that the free-induction
decay characterized by the evolution of optical coherences in the
coherence and echo times is well characterized, and given by expressions
of the form (\ref{eq:optical coherence}). The reason is that the
characteristic energetic scales of the vibrational degrees of freedom
are much lower than the optical gap, so the only nonunitary dynamics
they induce in the optical coherence is, to a good approximation,
restricted to pure dephasing $\Gamma_{ij}$ which can be inferred
from the polarization signal %
\footnote{We anticipate that deviations from $\mathcal{G}_{ij}(\tau)$, if they
were to happen, would most likely occur for short times $\tau$, where
the non-Markovian behavior of the bath will be stronger. This could
consist of a non-secular transfer of optical coherences, for instance:
$|g\rangle\langle\beta|\to|g\rangle\langle\alpha|$. However, as we
mention in section IV, the polarization $P_{PE}(\mbox{\ensuremath{\tau}},T,t)$
is collected for many $\tau$ and $t$ points, and subsequently Fourier
transformed along these dimensions. After processing the signal in
this way, the errors due to the short time deviations will presumably
be negligible, and the lineshape will be dominated by the $\mathcal{G}_{ij}(\tau)$
functional dependence. We note that these problems are not alien to
our protocol, but are generic concerns of any QPT with respect to
errors of in the preparation and measurement stages.%
}. The dynamics in the waiting time is more complex, consisting of
small frequencies due to excitonic superpositions which are strongly
influenced by the bath. It is the latter interval where QPT will prove
most useful.

The polarization signal yields a linear combination of elements $\chi_{abcd}(T)$
weighted by the probability amplitude to prepare a state $|c\rangle\langle d|$
with the first two pulses and detect $|a\rangle\langle b|$ with the
third pulse and the fourth heterodyning pulse. These probability amplitudes
can be controlled by manipulating the polarization of the pulses $\boldsymbol{e}_{i}$
and the frequency amplitudes for the resonant transitions $C_{\omega_{1}}^{p},C_{\omega_{2}}^{q},C_{\omega_{3}}^{r}$.
In essence, state preparation and QST are implicit in the coherence
and echo times (see \cite{yuenzhou}). In a different context, Gelin
and Kosov had previously hinted at a similar idea by identifying these
times as {}``doorway'' and {}``window'' intervals \cite{Gelin2008177}.
By conducting several experiments varying these control knobs and
collecting the signal from each of these settings, a system of linear
equations can be established whereby the elements of $\chi(T)$ can
be inverted, and therefore QPT is achieved. This statement is correct
provided that besides the free-induction decay rates $\Gamma_{ij}$,
the parameters $\omega_{\alpha g}$, $\omega_{\beta g}$, $\mathbb{\boldsymbol{\mu}}_{\alpha g}$,
$\mathbb{\boldsymbol{\mu}}_{\alpha g}$, $\mathbb{\boldsymbol{\mu}}_{\alpha g}$,
and $\mathbb{\boldsymbol{\mu}}_{\alpha g}$ are all known or can be
obtained self-consistently during the experiment. We will elaborate
on these points for the case of a homodimer in section V.

Notice that Eqs. (\ref{eq:total polarization}), (\ref{eq:with alpha}),
and (\ref{eq:with beta}) monitor all the 12 real valued paramenters
involving $\chi_{abcd}(T)$ for $a,b,c,d\in\{\alpha,\beta\}$, so
that they allow for the QPT of the single exciton manifold, which
is an effective quantum bit (qubit) system. However, we have also
kept track of the elements $\chi_{ggcd}(T)$ $c,d\in\{\alpha,\beta\}$,
that is, the possibility of amplitude leakage \emph{errors} from the
single-exciton dynamics channel to $|g\rangle\langle g|$. It is known
that whereas the excitonic dynamics occurs in femtosecond timescales,
exciton recombination happens in the order of nanoseconds. Therefore,
these decay channels could be potentially ignored in many experimental
systems. We shall keep them in our theoretical analysis as they do
not increase the complexity of the problem by much, although in situations
where this could be problematic, we could accordingly disregard them.

\section{QPT from 2D spectrum of PE}

As mentioned, QPT can be carried out from data resulting from a series
of experiments varying colors and polarizations of the pulses. The
necessary information can in principle be obtained by collecting a
single point for a fixed pair of $\tau$ and $t$ points for each
of the experiments. Often, however, the PE signal is collected across
many $\tau,T,t$ points, and conveniently processed into a 2D correlation
spectrum in the conjugate frequency variables $\omega_{\tau}$ and
$\omega_{t}$:

\begin{equation}
S(\omega_{\tau},T,\omega_{t})=i\int_{0}^{\infty}d\tau e^{-i\omega_{\tau}\tau}\int_{0}^{\infty}dte^{i\omega_{t}t}P_{PE}(\tau,T,t)\label{eq:integral}\end{equation}
which still evolves in the $T$ coordinate%
\footnote{The factor of $i$ arises due to the phase shift relating the macroscopic
polarization and the detected signal corresponding to the emitted
electric field.%
}. By performing the integrals of Eq. (\ref{eq:integral}) using Eqs.
(\ref{eq:total polarization}), (\ref{eq:with alpha}), and (\ref{eq:with beta}),
we obtain:

\begin{eqnarray}
S(\omega_{\tau},T,\omega_{t}) & = & i\sum_{m,n=\alpha,\beta}l_{\tau,m}(\omega_{\tau})l_{t,n}(\omega_{t})S_{mn}(T).\label{eq:2dspectrum expression}\end{eqnarray}
The spectrum consists of a sum of four resonances at $(\omega_{\tau},\omega_{t})\in\{(\omega_{\alpha g},\omega_{\alpha g}),(\omega_{\alpha g},\omega_{\beta g}),(\omega_{\beta g},\omega_{\alpha g}),(\omega_{\beta g},\omega_{\beta g})\}$,
which correspond to the frequencies of the optical coherences at the
coherence and echo times. These resonances are modulated by lineshape
functions of the form,

\begin{eqnarray}
l_{\tau,m}(\omega_{\tau}) & = & \frac{1}{i(\omega_{\tau}-\omega_{mg}-i\Gamma_{mg})},\label{eq:lineshape tau}\\
l_{t,n}(\omega_{t}) & = & \frac{1}{i(-\omega_{t}+\omega_{ng}-i\Gamma_{ng})},\label{eq:lineshape t}\end{eqnarray}
that correspond to the one-sided Fourier transform of the propagator
along each $\tau$ and $t$ axis%
\footnote{The fully dispersive and absorptive lineshapes only show up after
including the non-rephasing signal in the 2D-ES. See \cite{gallagher}
for more information on this issue.%
}. The peaks are centered about $\omega=\omega_{mg}$ and have a width
parameter $\Gamma_{mg}$. The difference in signs for the Fourier
transform in Eq. (\ref{eq:integral}) guarantees that all the resonances
appear in the first quadrant of both frequency axes. The expressions
for the amplitudes $S_{mn}(T)$, associated with peaks centered at
$(\omega_{\tau},\omega_{t})=(\omega_{mg},\omega_{ng})$, are given
by %
\footnote{These expressions were already displayed in \cite{procedia} without
the background in this article.%
}:

\begin{eqnarray}
S_{\alpha\alpha}(T) & = & -iC_{\omega_{1}}^{\alpha}C_{\omega_{2}}^{\alpha}(\boldsymbol{\mu}_{\alpha g}\cdot\boldsymbol{e}_{1})(\boldsymbol{\mu}_{\alpha g}\cdot\boldsymbol{e}_{2})\nonumber \\
 &  & \times\{C_{\omega_{3}}^{\alpha}[(\boldsymbol{\mu}_{\alpha g}\cdot\boldsymbol{e}_{3})(\boldsymbol{\mu}_{\alpha g}\cdot\boldsymbol{e}_{4})(\chi_{gg\alpha\alpha}(T)-1-\chi_{\alpha\alpha\alpha\alpha}(T))\nonumber \\
 &  & +(\boldsymbol{\mu}_{f\beta}\cdot\boldsymbol{e}_{3})(\boldsymbol{\mu}_{f\beta}\cdot\boldsymbol{e}_{4})\chi_{\beta\beta\alpha\alpha}(T)]\nonumber \\
 &  & +C_{\omega_{3}}^{\beta}[(\boldsymbol{\mu}_{f\alpha}\cdot\boldsymbol{e}_{3})(\boldsymbol{\mu}_{f\beta}\cdot\boldsymbol{e}_{4})-(\boldsymbol{\mu}_{\beta g}\cdot\boldsymbol{e}_{3})(\boldsymbol{\mu}_{\alpha g}\cdot\boldsymbol{e}_{4}))\chi_{\alpha\beta\alpha\alpha}(T)]\}\nonumber \\
 &  & -iC_{\omega_{1}}^{\alpha}C_{\omega_{2}}^{\beta}(\boldsymbol{\mu}_{\alpha g}\cdot\boldsymbol{e}_{1})(\boldsymbol{\mu}_{\beta g}\cdot\boldsymbol{e}_{2})\nonumber \\
 &  & \times\{C_{\omega_{3}}^{\alpha}[(\boldsymbol{\mu}_{\alpha g}\cdot\boldsymbol{e}_{3})(\boldsymbol{\mu}_{\alpha g}\cdot\boldsymbol{e}_{4})(\chi_{gg\beta\alpha}(T)-\chi_{\alpha\alpha\beta\alpha}(T))\nonumber \\
 &  & +(\boldsymbol{\mu}_{f\beta}\cdot\boldsymbol{e}_{3})(\boldsymbol{\mu}_{f\beta}\cdot\boldsymbol{e}_{4})\chi_{\beta\beta\beta\alpha}(T)]\nonumber \\
 &  & +C_{\omega_{3}}^{\beta}[((\boldsymbol{\mu}_{f\alpha}\cdot\boldsymbol{e}_{3})(\boldsymbol{\mu}_{f\beta}\cdot\boldsymbol{e}_{4})-(\boldsymbol{\mu}_{\beta g}\cdot\boldsymbol{e}_{3})(\boldsymbol{\mu}_{\alpha g}\cdot\boldsymbol{e}_{4}))\chi_{\alpha\beta\beta\alpha}(T)]\}\label{eq:saa}\end{eqnarray}

\begin{eqnarray}
S_{\alpha\beta}(T) & = & -iC_{\omega_{1}}^{\alpha}C_{\omega_{2}}^{\alpha}(\boldsymbol{\mu}_{\alpha g}\cdot\boldsymbol{e}_{1})(\boldsymbol{\mu}_{\alpha g}\cdot\boldsymbol{e}_{2})\nonumber \\
 &  & \times\{C_{\omega_{3}}^{\beta}[(\boldsymbol{\mu}_{\beta g}\cdot\boldsymbol{e}_{3})(\boldsymbol{\mu}_{\beta g}\cdot\boldsymbol{e}_{4})(\chi_{gg\alpha\alpha}(T)-1-\chi_{\beta\beta\alpha\alpha}(T))\nonumber \\
 &  & +(\boldsymbol{\mu}_{f\alpha}\cdot\boldsymbol{e}_{3})(\boldsymbol{\mu}_{f\alpha}\cdot\boldsymbol{e}_{4})\chi_{\alpha\alpha\alpha\alpha}(T)]\nonumber \\
 &  & +C_{\omega_{3}}^{\alpha}[((\boldsymbol{\mu}_{f\beta}\cdot\boldsymbol{e}_{3})(\boldsymbol{\mu}_{f\alpha}\cdot\boldsymbol{e}_{4})-(\boldsymbol{\mu}_{\alpha g}\cdot\boldsymbol{e}_{3})(\boldsymbol{\mu}_{\beta g}\cdot\boldsymbol{e}_{4}))\chi_{\beta\alpha\alpha\alpha}(T)]\}\nonumber \\
 &  & -iC_{\omega_{1}}^{\alpha}C_{\omega_{2}}^{\beta}(\boldsymbol{\mu}_{\alpha g}\cdot\boldsymbol{e}_{1})(\boldsymbol{\mu}_{\beta g}\cdot\boldsymbol{e}_{2})\nonumber \\
 &  & \times\{C_{\omega_{3}}^{\beta}[(\boldsymbol{\mu}_{\beta g}\cdot\boldsymbol{e}_{3})(\boldsymbol{\mu}_{\beta g}\cdot\boldsymbol{e}_{4})(\chi_{gg\beta\alpha}(T)-\chi_{\beta\beta\beta\alpha}(T))\nonumber \\
 &  & +(\boldsymbol{\mu}_{f\alpha}\cdot\boldsymbol{e}_{3})(\boldsymbol{\mu}_{f\alpha}\cdot\boldsymbol{e}_{4})\chi_{\alpha\alpha\beta\alpha}(T)]\nonumber \\
 &  & +C_{\omega_{3}}^{\alpha}[((\boldsymbol{\mu}_{f\beta}\cdot\boldsymbol{e}_{3})(\boldsymbol{\mu}_{f\alpha}\cdot\boldsymbol{e}_{4})-(\boldsymbol{\mu}_{\alpha g}\cdot\boldsymbol{e}_{3})(\boldsymbol{\mu}_{\beta g}\cdot\boldsymbol{e}_{4}))\chi_{\beta\alpha\beta\alpha}(T)]\}\label{eq:sab}\end{eqnarray}

\begin{eqnarray}
S_{\beta\beta}(T) & = & -iC_{\omega_{1}}^{\beta}C_{\omega_{2}}^{\beta}(\boldsymbol{\mu}_{\beta g}\cdot\boldsymbol{e}_{1})(\boldsymbol{\mu}_{\beta g}\cdot\boldsymbol{e}_{2})\nonumber \\
 &  & \times\{C_{\omega_{3}}^{\beta}[(\boldsymbol{\mu}_{\beta g}\cdot\boldsymbol{e}_{3})(\boldsymbol{\mu}_{\beta g}\cdot\boldsymbol{e}_{4})(\chi_{gg\beta\beta}(T)-1-\chi_{\beta\beta\beta\beta}(T))\nonumber \\
 &  & +(\boldsymbol{\mu}_{f\alpha}\cdot\boldsymbol{e}_{3})(\boldsymbol{\mu}_{f\alpha}\cdot\boldsymbol{e}_{4})\chi_{\alpha\alpha\beta\beta}(T)]\nonumber \\
 &  & +C_{\omega_{3}}^{\alpha}[((\boldsymbol{\mu}_{f\beta}\cdot\boldsymbol{e}_{3})(\boldsymbol{\mu}_{f\alpha}\cdot\boldsymbol{e}_{4})-(\boldsymbol{\mu}_{\alpha g}\cdot\boldsymbol{e}_{3})(\boldsymbol{\mu}_{\beta g}\cdot\boldsymbol{e}_{4}))\chi_{\beta\alpha\beta\beta}(T)]\}\nonumber \\
 &  & -iC_{\omega_{1}}^{\beta}C_{\omega_{2}}^{\alpha}(\boldsymbol{\mu}_{\beta g}\cdot\boldsymbol{e}_{1})(\boldsymbol{\mu}_{\alpha g}\cdot\boldsymbol{e}_{2})\nonumber \\
 &  & \times\{C_{\omega_{3}}^{\beta}[(\boldsymbol{\mu}_{\beta g}\cdot\boldsymbol{e}_{3})(\boldsymbol{\mu}_{\beta g}\cdot\boldsymbol{e}_{4})(\chi_{gg\alpha\beta}(T)-\chi_{\beta\beta\alpha\beta}(T))\nonumber \\
 &  & +(\boldsymbol{\mu}_{f\alpha}\cdot\boldsymbol{e}_{3})(\boldsymbol{\mu}_{f\alpha}\cdot\boldsymbol{e}_{4})\chi_{\alpha\alpha\alpha\beta}(T)]\nonumber \\
 &  & +C_{\omega_{3}}^{\alpha}[((\boldsymbol{\mu}_{f\beta}\cdot\boldsymbol{e}_{3})(\boldsymbol{\mu}_{f\alpha}\cdot\boldsymbol{e}_{4})-(\boldsymbol{\mu}_{\alpha g}\cdot\boldsymbol{e}_{3})(\boldsymbol{\mu}_{\beta g}\cdot\boldsymbol{e}_{4}))\chi_{\beta\alpha\alpha\beta}(T)]\}\label{eq:sbb}\end{eqnarray}

\begin{eqnarray}
S_{\beta\alpha}(T) & = & -iC_{\omega_{1}}^{\beta}C_{\omega_{2}}^{\beta}(\boldsymbol{\mu}_{\beta g}\cdot\boldsymbol{e}_{1})(\boldsymbol{\mu}_{\beta g}\cdot\boldsymbol{e}_{2})\nonumber \\
 &  & \times\{C_{\omega_{3}}^{\alpha}[(\boldsymbol{\mu}_{\alpha g}\cdot\boldsymbol{e}_{3})(\boldsymbol{\mu}_{\alpha g}\cdot\boldsymbol{e}_{4})(\chi_{gg\beta\beta}(T)-1-\chi_{\alpha\alpha\beta\beta}(T))\nonumber \\
 &  & +(\boldsymbol{\mu}_{f\beta}\cdot\boldsymbol{e}_{3})(\boldsymbol{\mu}_{f\beta}\cdot\boldsymbol{e}_{4})\chi_{\beta\beta\beta\beta}(T)]\nonumber \\
 &  & +C_{\omega_{3}}^{\beta}[(\boldsymbol{\mu}_{f\alpha}\cdot\boldsymbol{e}_{3})(\boldsymbol{\mu}_{f\beta}\cdot\boldsymbol{e}_{4})-(\boldsymbol{\mu}_{\beta g}\cdot\boldsymbol{e}_{3})(\boldsymbol{\mu}_{\alpha g}\cdot\boldsymbol{e}_{4}))\chi_{\alpha\beta\beta\beta}(T)]\}\nonumber \\
 &  & -iC_{\omega_{1}}^{\beta}C_{\omega_{2}}^{\alpha}(\boldsymbol{\mu}_{\beta g}\cdot\boldsymbol{e}_{1})(\boldsymbol{\mu}_{\alpha g}\cdot\boldsymbol{e}_{2})\nonumber \\
 &  & \times\{C_{\omega_{3}}^{\alpha}[(\boldsymbol{\mu}_{\alpha g}\cdot\boldsymbol{e}_{3})(\boldsymbol{\mu}_{\alpha g}\cdot\boldsymbol{e}_{4})(\chi_{gg\alpha\beta}(T)-\chi_{\alpha\alpha\alpha\beta}(T))\nonumber \\
 &  & +(\boldsymbol{\mu}_{f\beta}\cdot\boldsymbol{e}_{3})(\boldsymbol{\mu}_{f\beta}\cdot\boldsymbol{e}_{4})\chi_{\beta\beta\alpha\beta}(T)]\nonumber \\
 &  & +C_{\omega_{3}}^{\beta}[((\boldsymbol{\mu}_{f\alpha}\cdot\boldsymbol{e}_{3})(\boldsymbol{\mu}_{f\beta}\cdot\boldsymbol{e}_{4})-(\boldsymbol{\mu}_{\beta g}\cdot\boldsymbol{e}_{3})(\boldsymbol{\mu}_{\alpha g}\cdot\boldsymbol{e}_{4}))\chi_{\alpha\beta\alpha\beta}(T)]\}\label{eq:sba}\end{eqnarray}

Typically, the probed samples are in solution, so the molecules in
the ensemble are isotropically distributed. The isotropic average
$\langle\cdot\rangle$ for a tetradic $(\boldsymbol{\mu}_{a}\cdot\boldsymbol{e}_{1})(\boldsymbol{\mu}_{b}\cdot\boldsymbol{e}_{2})(\boldsymbol{\mu}_{c}\cdot\boldsymbol{e}_{3})(\boldsymbol{\mu}_{d}\cdot\boldsymbol{e}_{4})$
is given by \cite{thirunamachandran}:

\begin{eqnarray}
 &  & \langle(\boldsymbol{\mu}_{a}\cdot\boldsymbol{e}_{1})(\boldsymbol{\mu}_{b}\cdot\boldsymbol{e}_{2})(\boldsymbol{\mu}_{c}\cdot\boldsymbol{e}_{3})(\boldsymbol{\mu}_{d}\cdot\boldsymbol{e}_{4})\rangle\nonumber \\
 & = & \sum_{m_{1}m_{2}m_{3}m_{4}}I_{e_{1}e_{2}e_{3}e_{4};m_{1}m_{2}m_{3}m_{4}}^{(4)}\nonumber \\
 &  & \times[(\boldsymbol{\mu}_{a}\cdot\boldsymbol{m}_{1})(\boldsymbol{\mu}_{b}\cdot\boldsymbol{m}_{2})(\boldsymbol{\mu}_{c}\cdot\boldsymbol{m}_{3})(\boldsymbol{\mu}_{d}\cdot\boldsymbol{m}_{4})],\label{eq:isotropic tensor 1}\end{eqnarray}

\begin{eqnarray}
 &  & I_{e_{1}e_{2}e_{3}e_{4};m_{1}m_{2}m_{3}m_{4}}\nonumber \\
 & = & \frac{1}{30}\begin{array}{ccc}
[\delta_{e_{1}e_{2}}\delta_{e_{3}e_{4}} & \delta_{e_{1}e_{3}}\delta_{e_{2}e_{4}} & \delta_{e_{1}e_{4}}\delta_{e_{2}e_{3}}]\end{array}\nonumber \\
 &  & \times\left[\begin{array}{ccc}
4 & -1 & -1\\
-1 & 4 & -1\\
-1 & -1 & 4\end{array}\right]\left[\begin{array}{c}
\delta_{m_{1}m_{2}}\delta_{m_{3}m_{4}}\\
\delta_{m_{1}m_{3}}\delta_{m_{2}m_{4}}\\
\delta_{m_{1}m_{4}}\delta_{m_{2}m_{3}}\end{array}\right],\label{eq:isotropic tensor}\end{eqnarray}
where $\boldsymbol{e}_{i}$ and $\boldsymbol{m}_{i}$ are the polarizations
of the pulses in the lab and the molecular frame, respectively. The
isotropic average consists of a sum of molecular frame products $[(\boldsymbol{\mu}_{a}\cdot\boldsymbol{m}_{1})(\boldsymbol{\mu}_{b}\cdot\boldsymbol{m}_{2})(\boldsymbol{\mu}_{c}\cdot\boldsymbol{m}_{3})(\boldsymbol{\mu}_{d}\cdot\boldsymbol{m}_{4})]$
weighted by the isotropically invariant tensor $I_{e_{1}e_{2}e_{3}e_{4};m_{1}m_{2}m_{3}m_{4}}^{(4)}$. 

Since the information in Eqs. (\ref{eq:total polarization}), (\ref{eq:with alpha}),
(\ref{eq:with beta}) is in principle contained in Eqs. (\ref{eq:total polarization}),
(\ref{eq:with alpha}), and (\ref{eq:with beta}), several conclusions
from our previous study are immediately transferable: The elements
of $\chi(T)$ contained in Eqs. (\ref{eq:saa}), (\ref{eq:sab}),
(\ref{eq:sbb}), and (\ref{eq:sba}) can be all be extracted by repeating
a number of experiments with different polarization configurations
for the fields and two different waveforms for the pulses. Under different
motivations, theoretical proposals for manipulating 2D-ES using pulse-shaping
capabilities have been previously reported \cite{voronine:044508,abrchemphys}.
An extensive study of this possibility for a heterodimer will be presented
in the second article accompanying this study.

Eqs. (\ref{eq:saa}), (\ref{eq:sab}), (\ref{eq:sbb}), and (\ref{eq:sba})
can also be derived by book-keeping the double-sided Feynman diagrams
that oscillate at the particular frequencies for the coherence and
waiting times in each of the four resonances (we refer the reader
to Fig. 1). In analyzing the possible pathways in Liouville space,
we make use of the rotating-wave approximation (RWA): Perturbations
which are proportional to $e{}^{-i\boldsymbol{k}_{i}\cdot\boldsymbol{r}+i\omega_{i}(t-t_{i})}\hat{\boldsymbol{\mu}}\cdot\boldsymbol{e}_{i}$
can excite the ket and de-excite the bra, whereas the ones proportional
to $e{}^{i\boldsymbol{k}_{i}\cdot\boldsymbol{r}-i\omega_{i}(t-t_{i})}\hat{\boldsymbol{\mu}}\cdot\boldsymbol{e}_{i}$
can deexcite the ket and excite the bra. As an illustration, consider
the signal $S_{\alpha\beta}(T)$, which arises from diagrams oscillating
with frequency $\omega_{g\alpha}$ at the coherence time and $\omega_{\beta g}$
at the echo time (Fig. 1(b)). The two states at the coherence time
which can oscillate at $\omega_{g\alpha}$ are $|g\rangle\langle\alpha|$
or $|\beta\rangle\langle f|$, but the latter cannot be produced by
a single action of the dipole operator on the initial ground state
$|g\rangle\langle g|$. Therefore, $|g\rangle\langle\alpha|$ is the
only possible state for the coherence interval, and is produced by
acting the first pulse on the bra of the ground state: $|g\rangle\langle g|\to|g\rangle\langle\alpha|$.
Similar considerations imply that the state at the echo time must
be $|\beta\rangle\langle g|$ or $|f\rangle\langle\alpha|$. Given
these constraints, we are ready to enumerate the possible initial
and final states for the waiting time interval which are compatible
with these restrictions. By exciting the ket or deexciting the bra
of $|g\rangle\langle\alpha|$ with the second pulse, the following
initial states $|c\rangle\langle d|$ for the quantum channel can
be produced: $|c\rangle\langle d|\in\{|\alpha\rangle\langle\alpha|,|\beta\rangle\langle\alpha|,|g\rangle\langle g|\}$.
The final states $|a\rangle\langle b|\in\{|\alpha\rangle\langle\alpha|,|\beta\rangle\langle\alpha|,|\beta\rangle\langle\beta|,|g\rangle\langle g|\}$
can all give rise to $|\beta\rangle\langle g|$ or $|f\rangle\langle\alpha|$
by exciting the ket or deexciting the bra with the third pulse. Therefore,
in principle, there are $4\times3=12$ possibilities for $\chi_{abcd}(T)$
which can be detected in at $S_{\alpha\beta}(T)$. However, we assume
that the state $|g\rangle\langle g|$ does not evolve to other states
due to the bath:

\begin{equation}
\chi_{abgg}(T)=\delta_{ag}\delta_{bg},\label{eq:xgg}\end{equation}
This assumption is quite reasonable, as we are ignoring processes
where phonons can induce optical excitations from $|g\rangle\langle g|$.
This condition is present in Eqs. (\ref{eq:with alpha}), and (\ref{eq:with beta})
in the $\delta-$function terms and in Eqs. (\ref{eq:saa}), (\ref{eq:sab}),
(\ref{eq:sbb}), and (\ref{eq:sba}) in the {}``-1'' terms, which
correspond to $-\chi_{gggg}(T)$. This leaves $12-4=8$ possibilities
for $\chi_{abcd}(T)$. 

\begin{center}
\begin{figure}
\begin{centering}
\includegraphics[width=15cm]{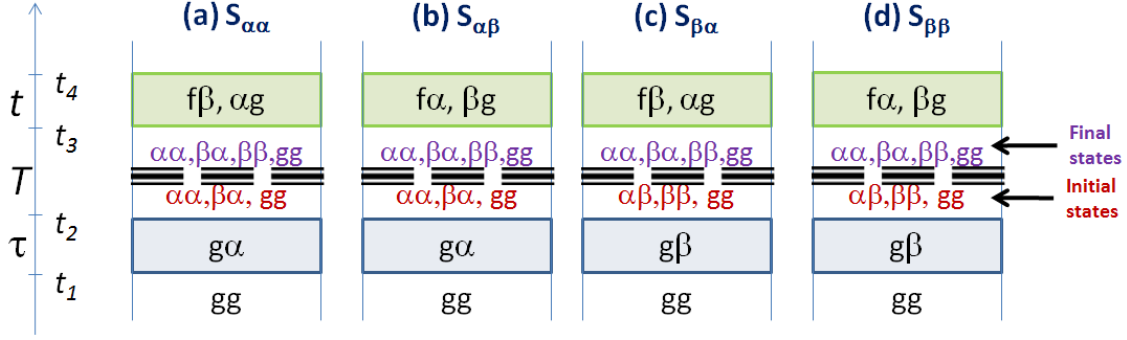} 
\par\end{centering}

\caption{Liouville space pathways corresponding to each of the four resonances
in the 2D-ES of a coupled dimer. The amplitude $S_{mn}(T)$ corresponds
to the peak located at $(\omega_{\tau},\omega_{t})=(\omega_{mg},\omega_{ng})$,
which are the values of the optical frequencies at the coherence and
echo time intervals $\tau$ and $t$, respectively. These amplitudes
provide information on the coherent and incoherent excitonic processes
at the waiting time $T$ by enumerating the possible initial and final
states at the waiting time $T$ which satisfy the PE phase matching
condition for the pulses acting at times $t_{1},t_{2},t_{3},t_{4}$.
The information contained in the amplitudes $S_{mn}(T)$ can be distilled
to reconstruct the process matrix for the single-exciton manifold
of the dimer, thus allowing for a QPT.}

\end{figure}

\par\end{center}

To be more explicit, consider the pathways in $S_{\alpha\beta}(T)$
that monitor the population to coherence process $\chi_{\beta\alpha\alpha\alpha}(T)$.
These are displayed in Fig. (2). The pathway on the left involves
represents an excited state absorption (ESA) from the single-exciton
manifold, and is proportional to $(-C_{\omega_{1}}^{\alpha}\boldsymbol{\mu}_{\alpha g}\cdot\boldsymbol{e}_{1})(C_{\omega_{2}}^{\alpha}\boldsymbol{\mu}_{\alpha g}\cdot\boldsymbol{e}_{2})(C_{\omega_{3}}^{\alpha}\boldsymbol{\mu}_{f\beta}\cdot\boldsymbol{e}_{3})(\boldsymbol{\mu}_{f\alpha}\cdot\boldsymbol{e}_{4})$,
an expression which can be immediately read out from the diagram:
Each interaction with the field picks up a factor corresponding to
the amplitude of the transition, which depends on the alignment of
the corresponding dipole with the polarization of the pulse, as well
as the frequency amplitude of the pulse at the given transition. A
minus sign is included if the perturbation is on the bra. Similarly,
the pathway on the right involves stimulated emission (SE) and is
proportional to $(-C_{\omega_{1}}^{\alpha}\boldsymbol{\mu}_{\alpha g}\cdot\boldsymbol{e}_{1})(C_{\omega_{2}}^{\alpha}\boldsymbol{\mu}_{\alpha g}\cdot\boldsymbol{e}_{2})(-C_{\omega_{3}}^{\alpha}\boldsymbol{\mu}_{\alpha g}\cdot\boldsymbol{e}_{3})(\boldsymbol{\mu}_{\beta g}\cdot\boldsymbol{e}_{4})$. 

The rest of the diagrams for all the peaks can be systematically analyzed
in the way described above. In general, the pathways we need to consider
can be classified in ESA, SE, and ground state bleaching (GSB) processes.
GSB processes are the ones that take $|g\rangle\langle g|$ at the
end of the waiting time to a dipole active coherence involving an
excited state. ESA pathways, which are proportional to dipole transitions
involving the excited state, differ in sign from SE and GSB pathways,
as can be easily seen by inspection. 

\begin{center}
\begin{figure}
\begin{centering}
\includegraphics[width=15cm]{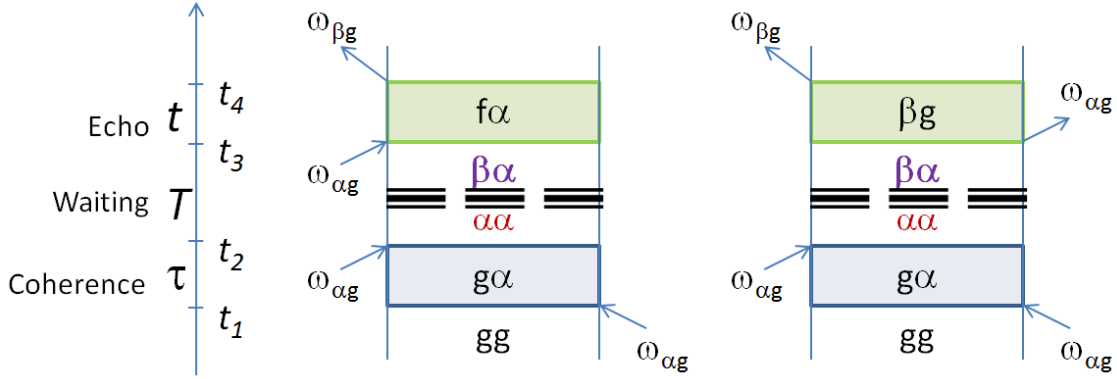} 
\par\end{centering}

\caption{A more detailed view on the Liouville space pathways corresponding
to the monitoring of the population to coherence process $\chi_{\beta\alpha\alpha\alpha}(T)$
in the peak at $(\omega_{\tau},\omega_{t})=(\omega_{\alpha g},\omega_{\beta g})$
of the 2D-ES. These diagrams belong to the amplitude $S_{\alpha\beta}(T)$
and can be easy constructed by taking into account the PE phase-matching
and the resonant conditions. }

\end{figure}

\par\end{center}

Fig. (3) provides a mnemonic device to keep track of the Liouville
pathways that each peak in the 2D electronic spectrum monitors, and
therefore, also provides a scheme of the QPT protocol. The $\omega_{\tau}$
axis can be associated with a particular state preparation whereas
the $\omega_{t}$ axis with a particular detection. Each peak reflects
a nontrivial number of processes in Liouville space. As an illustration
(see Fig. (4)), we consider the ideal case where the bath does not
interact with the system, in which case, a very simple picture is
recovered: The off-diagonal peaks beat at the coherence frequency
and the diagonals remain static. This case can be easily derived from
Eqs. (\ref{eq:saa}), (\ref{eq:sab}), (\ref{eq:sbb}), and (\ref{eq:sba})
by substituting $\chi_{abcd}(T)=\delta_{ab}\delta_{cd}+\delta_{ac}\delta_{bd}e^{-i\omega_{ab}T}$,
that is, populations remain static whereas coherences beat at difference
frequencies.

\begin{center}
\begin{figure}
\begin{centering}
\includegraphics[width=8.5cm]{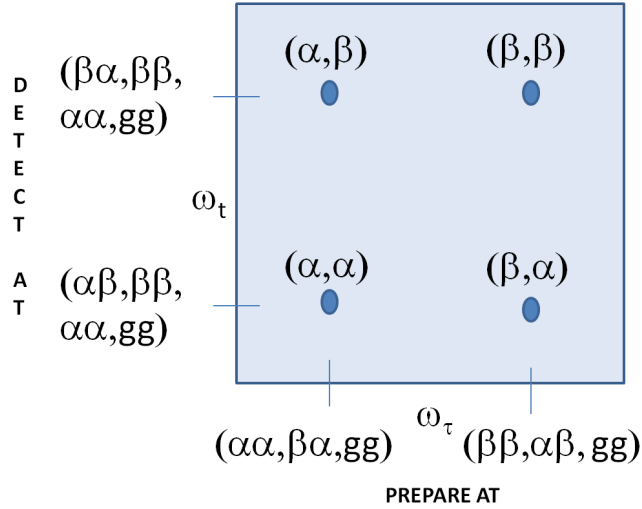} 
\par\end{centering}

\caption{Summary of QPT for a coupled dimer in the 2D-ES. The Liouville pathways
depicted in Fig. 1 can be condensed into this diagram. The horizontal
axis for the coherence frequency $\omega_{\tau}$ is associated with
a state preparation, whereas the vertical axis for the echo frequency
$\omega_{t}$ corresponds to a detection. The four resonances labeled
as $(m,n)$ correspond to peaks located at $(\omega_{\tau},\omega_{t})=(\omega_{mg},\omega_{ng})$.
Their amplitudes contain information on $\chi_{abcd}(T)$, where $cd$
is the state prepared at the beginning of the waiting time interval,
and $ab$ the state detected at the end of the same interval. For
instance, the peak at $(\omega_{\alpha},\omega_{\beta})$ keeps track
of the elements $\chi_{abcd}(T)$ where $|c\rangle\langle d|\in\{|\alpha\rangle\langle\alpha|,|\beta\rangle\langle\alpha|,|g\rangle\langle g|\}$
and $|a\rangle\langle b|\in\{|\beta\rangle\langle\alpha|,|\beta\rangle\langle\beta|,|\alpha\rangle\langle\alpha|,|g\rangle\langle g|\}$.}

\end{figure}

\par\end{center}

\begin{center}
\begin{figure}
\begin{centering}
\includegraphics[width=8.5cm]{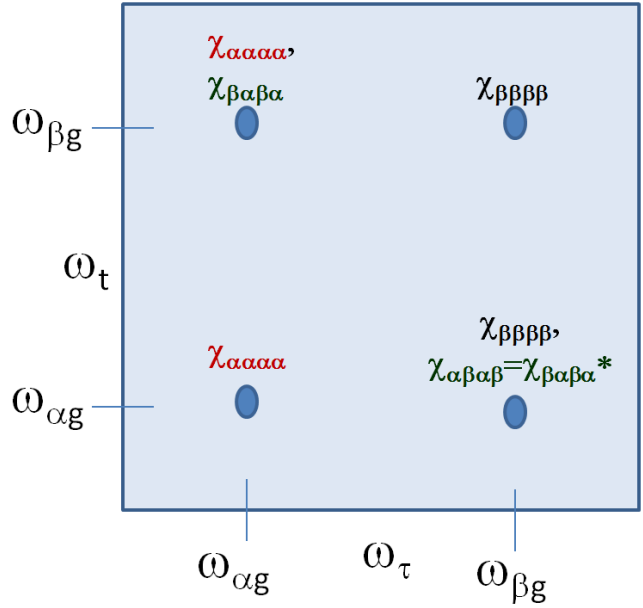} 
\par\end{centering}

\caption{2D-ES of a coupled dimer in the absence of interactions with a bath.
Under unitary dynamics of the excitonic system, each of the four resonances
keep track of the elements of $\chi(T)$ indicated in the diagram.
Notice that diagonal peaks do not oscillate as a function of waiting
time $T$, whereas off-diagonals beat at the frequency equal to the
difference in energies of the single exciton eigenstates. }

\end{figure}

\par\end{center}

\section{The case of the homodimer}

To gain insights into the described QPT protocol, we specialize the
results above to a coupled homodimer. In the following subsections,
we discuss, for this particular case, (A) the Hamiltonian and the
transition dipole moments involved in the experiments, (B) properties
of the spectroscopic signals under isotropic averaging, (C) stability
of the numerical inversion, (D) analytical expressions of the elements
of $\chi(T)$ in terms of the peak amplitudes of the spectra, (E)
a procedure to extract the angle $\phi$ between the dipoles, (F)
a summary of the QPT procedure, and (G) a numerical example with a
model system. A similar study focused on the heterodimer will be presented
in the second part of this series.

\subsection{Hamiltonian and transition dipole moments}

In the homodimer, the two sites are identical chromophores with energies
$\omega_{A}=\omega_{B}=\bar{\omega}$, and the Hamiltonian in Eq.
(\ref{eq:excitonic H not diagonalized-1}) and (\ref{eq:excitonic H-1})
is given by:

\begin{eqnarray}
H_{S} & = & \bar{\omega}(a_{A}^{+}a_{A}+a_{B}^{+}a_{B})+J(a_{A}^{+}a_{B}+a_{B}^{+}a_{A})\nonumber \\
 & = & (\bar{\omega}-J)a_{\alpha}^{+}a_{\alpha}+(\bar{\omega}+J)a_{\beta}^{+}a_{\beta},\label{eq:homodimer hamiltonian}\end{eqnarray}
which we have diagonalized with the symmetric $a_{\beta}^{+}|g\rangle$
and antisymmetric $a_{\alpha}^{+}|g\rangle$ single exciton states
given by:

\begin{eqnarray}
a_{\alpha}^{+} & = & \frac{1}{\sqrt{2}}(a_{A}^{+}+a_{B}^{+}),\nonumber \\
a_{\beta}^{+} & = & \frac{1}{\sqrt{2}}(a_{A}^{+}-a_{B}^{+}).\label{eq:creation homodimer}\end{eqnarray}
The splitting between the two delocalized states is just $2J$. Using
Eq. (\ref{eq:dipoles}), the transition dipoles take the simple forms:

\begin{eqnarray}
\boldsymbol{\mu}_{\alpha g} & = & \frac{1}{\sqrt{2}}(\boldsymbol{d}_{A}+\boldsymbol{d}_{B}),\nonumber \\
\boldsymbol{\mu}_{\beta g} & = & \frac{1}{\sqrt{2}}(\boldsymbol{d}_{A}-\boldsymbol{d}_{B}),\nonumber \\
\boldsymbol{\mu}_{f\alpha} & = & \boldsymbol{\mu}_{\alpha g},\nonumber \\
\boldsymbol{\mu}_{f\beta} & = & -\boldsymbol{\mu}_{\beta g}.\label{eq:dipoles homodimer-2}\end{eqnarray}
Interestingly, these expressions are independent of the coupling $J$.
Also, notice that $\boldsymbol{\mu}_{\alpha g}$ and $\boldsymbol{\mu}_{f\alpha}$
are perpendicular to $\boldsymbol{\mu}_{\beta g}$ and $\boldsymbol{\mu}_{f\beta}$
(see Fig. 5). Denoting the norm of each site dipole by \begin{eqnarray}
|\boldsymbol{d}_{A}| & = & |\boldsymbol{d}_{B}|=d,\label{eq:dipole equals d}\end{eqnarray}
the following relationships follow: 

\begin{eqnarray}
|\boldsymbol{\mu}_{\alpha g}| & = & |\boldsymbol{\mu}_{f\alpha}|=\mu_{\alpha g}=\sqrt{2}d\cos\left(\frac{\phi}{2}\right),\nonumber \\
|\boldsymbol{\mu}_{\beta g}| & = & |\boldsymbol{\mu}_{f\beta}|=\mu_{\beta g}=\sqrt{2}d\sin\left(\frac{\phi}{2}\right).\label{eq:dipoles homodimer-1}\end{eqnarray}

As expected, in the degenerate limit that $\phi=0$ or $\pi$ (the
site dipoles are parallel or antiparallel), one of the delocalized
excitons becomes dark and there is only one bright transition from
the ground state. If this is not the case, in general, the two transitions
are bright and their dipoles perpendicular to each other. Furthermore,
as a difference with the heterodimer case, there are only three (instead
of four) different transition dipoles in the homodimer, and two of
them are just negatives of each other. The degenerate case will be
discussed as a limit of the more general one in the next paragraphs.

\begin{center}
\begin{figure}
\begin{centering}
\includegraphics[width=15cm]{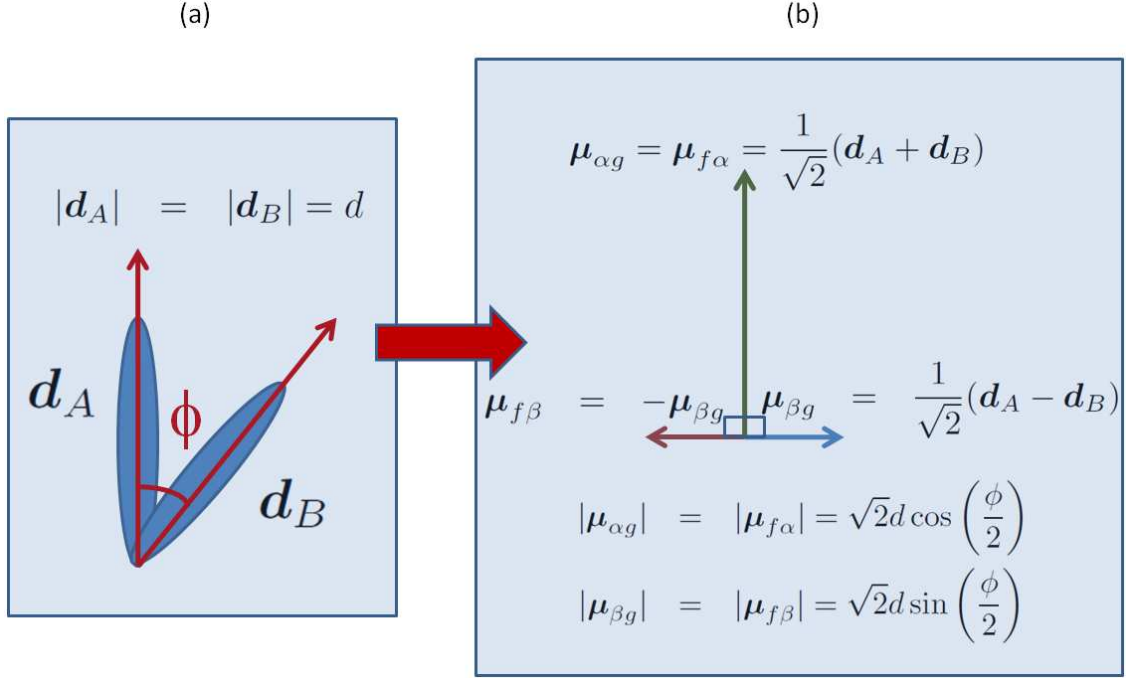} 
\par\end{centering}

\caption{Transition dipole moments of a homodimer. Diagrams for (a) sites and
(b) eigenstates.}

\end{figure}

\par\end{center}

\subsection{Isotropic averaging of signals}

An important observation regarding isotropic averaging follows:

\emph{\noun{Claim.--}} Upon isotropic averaging, signals stemming
from coherence to population or population to coherence transfer cannot
be monitored in the 2D-PE spectrum of a homodimer. 

\emph{\noun{Proof.--}} For simplicity, align $\boldsymbol{\mu}_{\beta g}$
and $\boldsymbol{\mu}_{\alpha g}$ in the $\boldsymbol{y}$ and $\boldsymbol{z}$
directions in the frame of the molecule and use Eq. (\ref{eq:isotropic tensor 1})
to argue that the only terms in the sum that contribute to an isotropic
averaging are the ones where only two or four polarizations of the
field are the same. In Eqs. (\ref{eq:saa}), (\ref{eq:sab}), (\ref{eq:sbb}),
and (\ref{eq:sba}), all the dipole-polarization terms corresponding
to coherence to population and the opposite processes involve three
dipoles of the same kind and a perpendicular one. Therefore, they
vanish under isotropic averaging. As an example, consider the the
terms associated with $\chi_{\beta\alpha\alpha\alpha}(T)$ in $S_{\alpha\beta}(T)$:

\begin{eqnarray*}
 &  & \langle(-C_{\omega_{1}}^{\alpha}\boldsymbol{\mu}_{\alpha g}\cdot\boldsymbol{e}_{1})(C_{\omega_{2}}^{\alpha}\boldsymbol{\mu}_{\alpha g}\cdot\boldsymbol{e}_{2})(C_{\omega_{3}}^{\alpha}\boldsymbol{\mu}_{f\beta}\cdot\boldsymbol{e}_{3})(\boldsymbol{\mu}_{f\alpha}\cdot\boldsymbol{e}_{4})\rangle\\
 & \propto & \langle(\boldsymbol{\mu}_{\alpha g}\cdot\boldsymbol{e}_{1})(\boldsymbol{\mu}_{\alpha g}\cdot\boldsymbol{e}_{2})(\boldsymbol{\mu}_{\beta g}\cdot\boldsymbol{e}_{3})(\boldsymbol{\mu}_{\alpha g}\cdot\boldsymbol{e}_{4})\rangle_{iso}\\
 & = & 0,\\
 &  & (\langle-C_{\omega_{1}}^{\alpha}\boldsymbol{\mu}_{\alpha g}\cdot\boldsymbol{e}_{1})(C_{\omega_{2}}^{\alpha}\boldsymbol{\mu}_{\alpha g}\cdot\boldsymbol{e}_{2})(-C_{\omega_{3}}^{\alpha}\boldsymbol{\mu}_{\alpha g}\cdot\boldsymbol{e}_{3})(\boldsymbol{\mu}_{\beta g}\cdot\boldsymbol{e}_{4})\rangle\\
 & \propto & \langle(\boldsymbol{\mu}_{\alpha g}\cdot\boldsymbol{e}_{1})(\boldsymbol{\mu}_{\alpha g}\cdot\boldsymbol{e}_{2})(\boldsymbol{\mu}_{\alpha g}\cdot\boldsymbol{e}_{3})(\boldsymbol{\mu}_{\beta g}\cdot\boldsymbol{e}_{4})\rangle\\
 & = & 0.\end{eqnarray*}

\begin{flushright}
$\Box$
\par\end{flushright}

The claim above allows for a considerable simplification of Eqs. (\ref{eq:saa}),
(\ref{eq:sab}), (\ref{eq:sbb}), and (\ref{eq:sba}). Each of the
peaks in the 2D spectrum keeps track of less elements of the process
matrix $\chi(T)$ upon isotropic averaging: Only population to population
and coherence to coherence transfers can be monitored. The results
for the peaks of an isotropically averaged 2D spectra are presented
below. We have taken the shortcut notation $\langle\cdot\rangle_{e_{1}e_{2}e_{3}e_{4}}$,
which denotes the isotropically averaged signal stemming from the
two pulse polarization configurations $(\boldsymbol{e}_{1},\boldsymbol{e}_{2},\boldsymbol{e}_{3},\boldsymbol{e}_{4})=(\boldsymbol{z},\boldsymbol{z},\boldsymbol{z},\boldsymbol{z}),(\boldsymbol{z},\boldsymbol{z},\boldsymbol{x},\boldsymbol{x})$,
so that the terms $\langle S_{mn}(T)\rangle_{e_{1}e_{2}e_{3}e_{4}}$
and $\langle S(\omega_{\tau},T,\omega_{t})\rangle_{e_{1}e_{2}e_{3}e_{4}}$
have the obvious meanings.

\begin{center}
\begin{tabular}{cc}
\hline 
\multicolumn{2}{c}{TABLE 1. Isotropically averaged 2D-ES peak amplitudes for the $\boldsymbol{z}\boldsymbol{z}\boldsymbol{z}\boldsymbol{z}$
configuration}\tabularnewline
\hline
{\footnotesize }%
\begin{minipage}[t]{0.5\columnwidth}%
{\footnotesize \begin{eqnarray*}
 &  & \langle S_{\alpha\beta}(T)\rangle_{zzzz}\\
 & = & -iC_{\omega_{1}}^{\alpha}C_{\omega_{2}}^{\alpha}C_{\omega_{3}}^{\beta}\\
 &  & \times[\frac{1}{15}\mu_{\alpha g}^{2}\mu_{\beta g}^{2}(\chi_{gg\alpha\alpha}(T)-1-\chi_{\beta\beta\alpha\alpha}(T))\\
 &  & +\frac{1}{5}\mu_{\alpha g}^{4}\chi_{\alpha\alpha\alpha\alpha}(T)]\\
 &  & -iC_{\omega_{1}}^{\alpha}C_{\omega_{2}}^{\beta}C_{\omega_{3}}^{\alpha}[(-)\frac{2}{15}\mu_{\alpha g}^{2}\mu_{\beta g}^{2}\chi_{\beta\alpha\beta\alpha}(T)]\end{eqnarray*}
}%
\end{minipage} & {\footnotesize }%
\begin{minipage}[t]{0.5\textwidth}%
{\footnotesize \begin{eqnarray*}
 &  & \langle S_{\beta\beta}(T)\rangle_{zzzz}\\
 & = & -iC_{\omega_{1}}^{\beta}C_{\omega_{2}}^{\beta}C_{\omega_{3}}^{\beta}\\
 &  & \times[\frac{1}{5}\mu_{\beta g}^{4}(\chi_{gg\beta\beta}(T)-1-\chi_{\beta\beta\beta\beta}(T))\\
 &  & +\frac{1}{15}\mu_{\beta g}^{2}\mu_{\alpha g}^{2}\chi_{\alpha\alpha\beta\beta}(T)]\\
 &  & -iC_{\omega_{1}}^{\beta}C_{\omega_{2}}^{\alpha}C_{\omega_{3}}^{\alpha}[(-)\frac{2}{15}\mu_{\beta g}^{2}\mu_{\alpha g}^{2}\chi_{\beta\alpha\alpha\beta}(T)]\end{eqnarray*}
}%
\end{minipage}\tabularnewline
{\footnotesize }%
\begin{minipage}[t]{0.5\columnwidth}%
{\footnotesize \begin{eqnarray*}
 &  & \langle S_{\alpha\alpha}(T)\rangle_{zzzz}\\
 & = & -iC_{\omega_{1}}^{\alpha}C_{\omega_{2}}^{\alpha}C_{\omega_{3}}^{\alpha}\\
 &  & \times[\frac{1}{5}\mu_{\alpha g}^{4}(\chi_{gg\alpha\alpha}(T)-1-\chi_{\alpha\alpha\alpha\alpha}(T))\\
 &  & +\frac{1}{15}\mu_{\alpha g}^{2}\mu_{\beta g}^{2}\chi_{\beta\beta\alpha\alpha}(T)]\\
 &  & -iC_{\omega_{1}}^{\alpha}C_{\omega_{2}}^{\beta}C_{\omega_{3}}^{\beta}[(-)\frac{2}{15}\mu_{\alpha g}^{2}\mu_{\beta g}^{2}\chi_{\alpha\beta\beta\alpha}(T)]\end{eqnarray*}
}%
\end{minipage} & {\footnotesize }%
\begin{minipage}[t]{0.5\columnwidth}%
{\footnotesize \begin{eqnarray*}
 &  & \langle S_{\beta\alpha}(T)\rangle_{zzzz}\\
 & = & -iC_{\omega_{1}}^{\beta}C_{\omega_{2}}^{\beta}C_{\omega_{3}}^{\alpha}\\
 &  & \times[\frac{1}{15}\mu_{\beta g}^{2}\mu_{\alpha g}^{2}(\chi_{gg\beta\beta}(T)-1-\chi_{\alpha\alpha\beta\beta}(T))\\
 &  & +\frac{1}{5}\mu_{\beta g}^{4}\chi_{\beta\beta\beta\beta}(T)]\\
 &  & -iC_{\omega_{1}}^{\beta}C_{\omega_{2}}^{\alpha}C_{\omega_{3}}^{\beta}[(-)\frac{2}{15}\mu_{\beta g}^{2}\mu_{\alpha g}^{2}\chi_{\alpha\beta\alpha\beta}(T)]\end{eqnarray*}
}%
\end{minipage}\tabularnewline
\hline
\end{tabular}
\par\end{center}

\pagebreak{}

\begin{center}
\begin{tabular}{cc}
\hline 
\multicolumn{2}{c}{TABLE 2. Isotropically averaged 2D-ES peak amplitudes for the $\boldsymbol{z}\boldsymbol{z}\boldsymbol{x}\boldsymbol{x}$
configuration }\tabularnewline
\hline
{\small }%
\begin{minipage}[t]{0.5\columnwidth}%
{\small \begin{eqnarray*}
 &  & \langle S{}_{\alpha\beta}(T)\rangle_{zzxx}\\
 & = & -iC_{\omega_{1}}^{\alpha}C_{\omega_{2}}^{\alpha}C_{\omega_{3}}^{\beta}\\
 &  & \times[\frac{2}{15}\mu_{\alpha g}^{2}\mu_{\beta g}^{2}(\chi_{gg\alpha\alpha}(T)-1-\chi_{\beta\beta\alpha\alpha}(T))\\
 &  & +\frac{1}{15}\mu_{\alpha g}^{4}\chi_{\alpha\alpha\alpha\alpha}(T)]\\
 &  & -iC_{\omega_{1}}^{\alpha}C_{\omega_{2}}^{\beta}C_{\omega_{3}}^{\alpha}[(+)\frac{1}{15}\mu_{\alpha g}^{2}\mu_{\beta g}^{2}\chi_{\beta\alpha\beta\alpha}(T)]\end{eqnarray*}
}%
\end{minipage} & {\small }%
\begin{minipage}[t]{0.5\columnwidth}%
{\small \begin{eqnarray*}
 &  & \langle S_{\beta\beta}(T)\rangle_{zzxx}\\
 & = & -iC_{\omega_{1}}^{\beta}C_{\omega_{2}}^{\beta}C_{\omega_{3}}^{\beta}\\
 &  & \times[\frac{1}{15}\mu_{\beta g}^{4}(\chi_{gg\beta\beta}(T)-1-\chi_{\beta\beta\beta\beta}(T))\\
 &  & +\frac{2}{15}\mu_{\beta g}^{2}\mu_{\alpha g}^{2}\chi_{\alpha\alpha\beta\beta}(T)]\\
 &  & -iC_{\omega_{1}}^{\beta}C_{\omega_{2}}^{\alpha}C_{\omega_{3}}^{\alpha}[(+)\frac{1}{15}\mu_{\beta g}^{2}\mu_{\alpha g}^{2}\chi_{\beta\alpha\alpha\beta}(T)]\end{eqnarray*}
}%
\end{minipage}\tabularnewline
{\small }%
\begin{minipage}[t]{0.5\columnwidth}%
{\small \begin{eqnarray*}
 &  & \langle S_{\alpha\alpha}(T)\rangle_{zzxx}\\
 & = & -iC_{\omega_{1}}^{\alpha}C_{\omega_{2}}^{\alpha}C_{\omega_{3}}^{\alpha}\\
 &  & \times[\frac{1}{15}\mu_{\alpha g}^{4}(\chi_{gg\alpha\alpha}(T)-1-\chi_{\alpha\alpha\alpha\alpha}(T))\\
 &  & +\frac{2}{15}\mu_{\alpha g}^{2}\mu_{\beta g}^{2}\chi_{\beta\beta\alpha\alpha}(T)]\\
 &  & -iC_{\omega_{1}}^{\alpha}C_{\omega_{2}}^{\beta}C_{\omega_{3}}^{\beta}[(+)\frac{1}{15}\mu_{\alpha g}^{2}\mu_{\beta g}^{2}\chi_{\alpha\beta\beta\alpha}(T)]\end{eqnarray*}
}%
\end{minipage} & {\small }%
\begin{minipage}[t]{0.5\columnwidth}%
{\small \begin{eqnarray*}
 &  & \langle S_{\beta\alpha}(T)\rangle_{zzxx}\\
 & = & -iC_{\omega_{1}}^{\beta}C_{\omega_{2}}^{\beta}C_{\omega_{3}}^{\alpha}\\
 &  & \times[\frac{2}{15}\mu_{\beta g}^{2}\mu_{\alpha g}^{2}(\chi_{gg\beta\beta}(T)-1-\chi_{\alpha\alpha\beta\beta}(T))\\
 &  & +\frac{1}{15}\mu_{\beta g}^{4}\chi_{\beta\beta\beta\beta}(T)]\\
 &  & -iC_{\omega_{1}}^{\beta}C_{\omega_{2}}^{\alpha}C_{\omega_{3}}^{\beta}[(+)\frac{1}{15}\mu_{\beta g}^{2}\mu_{\alpha g}^{2}\chi_{\alpha\beta\alpha\beta}(T)]\end{eqnarray*}
}%
\end{minipage}\tabularnewline
\hline
\end{tabular}
\par\end{center}

\pagebreak{}

We focus our attention on experiments with short pulses that are broadband
enough to create either exciton $|\alpha\rangle$ or $|\beta\rangle$
with the same amplitude, that is, $C_{\omega_{i}}^{p}=C$, for a purely
imaginary constant $C$, for all $p$ and $\omega_{i}$. This condition
can be easily relaxed, but we shall proceed with it to analyze our
QPT protocol with more detail. Using the condition of Eq. (\ref{eq:trace}),
we can eliminate the variables $\chi_{gg\alpha\alpha}(T)$ and $\chi_{gg\beta\beta}(T)$
for $\chi_{\alpha\alpha\alpha\alpha}(T),\chi_{\beta\beta\alpha\alpha}(T),\chi_{\alpha\alpha\beta\beta}(T),\chi_{\beta\beta\beta\beta}(T)$.
Also, taking advantage of Eq. (\ref{eq:hermiticity-1}), we discard
$\chi_{\beta\alpha\beta\alpha}(T)$ and $\chi_{\alpha\beta\beta\alpha}(T)$
and keep $\chi_{\alpha\beta\alpha\beta}(T)$ and $\chi_{\beta\alpha\alpha\beta}(T)$.
From the left hand side (LHS) of the real and imaginary parts of the
spectrum (see Tables 1 and 2), we derive the following real valued
matrix equation:

{\tiny \begin{eqnarray}
iC^{3}\left[\begin{array}{cccccc}
\frac{2}{5}\mu_{\alpha g}^{4} & \frac{1}{5}\mu_{\alpha g}^{4}-\frac{1}{15}\mu_{\alpha g}^{2}\mu_{\beta g}^{2} & 0 & \frac{2}{15}\mu_{\alpha g}^{2}\mu_{\beta g}^{2} & 0 & 0\\
\frac{2}{15}\mu_{\alpha g}^{4} & \frac{1}{15}\mu_{\alpha g}^{4}-\frac{2}{15}\mu_{\alpha g}\mu_{\beta g}^{2} & 0 & -\frac{1}{15}\mu_{\alpha g}^{2}\mu_{\beta g}^{2} & 0 & 0\\
-\frac{1}{5}\mu_{\alpha g}^{4}+\frac{1}{15}\mu_{\alpha g}^{2}\mu_{\beta g}^{2} & \frac{2}{15}\mu_{\alpha g}^{2}\mu_{\beta g}^{2} & \frac{2}{15}\mu_{\alpha g}^{2}\mu_{\beta g}^{2} & 0 & 0 & 0\\
-\frac{1}{15}\mu_{\alpha g}^{4}+\frac{2}{15}\mu_{\alpha g}^{2}\mu_{\beta g}^{2} & \frac{4}{15}\mu_{\alpha g}^{2}\mu_{\beta g}^{2} & -\frac{1}{15}\mu_{\alpha g}^{2}\mu_{\beta g}^{2} & 0 & 0 & 0\\
0 & 0 & 0 & 0 & 0 & -\frac{2}{15}\mu_{\alpha g}^{2}\mu_{\beta g}^{2}\\
0 & 0 & 0 & 0 & 0 & \frac{1}{15}\mu_{\alpha g}^{2}\mu_{\beta g}^{2}\\
0 & 0 & 0 & 0 & -\frac{2}{15}\mu_{\alpha g}^{2}\mu_{\beta g}^{2} & 0\\
0 & 0 & 0 & 0 & \frac{1}{15}\mu_{\alpha g}^{2}\mu_{\beta g}^{2} & 0\end{array}\right]\left[\begin{array}{c}
\chi_{\alpha\alpha\alpha\alpha}(T)\\
\chi_{\beta\beta\alpha\alpha}(T)\\
\Re\{\chi_{\alpha\beta\alpha\beta}(T)\}\\
\Re\{\chi_{\beta\alpha\alpha\beta}(T)\}\\
\Im\{\chi_{\alpha\beta\alpha\beta}(T)\}\\
\Im\{\chi_{\beta\alpha\alpha\beta}(T)\}\end{array}\right] & = & \left[\begin{array}{c}
\Re\{\langle S_{\alpha\alpha}(T)\rangle_{zzzz}\}\\
\Re\{\langle S_{\alpha\alpha}(T)\rangle_{zzxx}\}\\
\Re\{\langle S_{\alpha\beta}(T)\rangle_{zzzz}\}\\
\Re\{\langle S_{\alpha\beta}(T)\rangle_{zzxx}\}\\
\Im\{\langle S_{\alpha\alpha}(T)\rangle_{zzzz}\}\\
\Im\{\langle S_{\alpha\alpha}(T)\rangle_{zzxx}\}\\
\Im\{\langle S_{\alpha\beta}(T)\rangle_{zzzz}\}\\
\Im\{\langle S_{\alpha\beta}(T)\rangle_{zzxx}\}\end{array}\right].\label{eq:leftside}\end{eqnarray}
}{\tiny \par}

Similarly, the right hand side (RHS) of the spectra yields:

{\tiny \begin{eqnarray}
iC^{3}\left[\begin{array}{cccccc}
\frac{2}{5}\mu_{\beta g}^{4} & \frac{1}{5}\mu_{\beta g}^{4}-\frac{1}{15}\mu_{\beta g}^{2}\mu_{\alpha g}^{2} & 0 & \frac{2}{15}\mu_{\beta g}^{2}\mu_{\alpha g}^{2} & 0 & 0\\
\frac{2}{15}\mu_{\beta g}^{4} & \frac{1}{15}\mu_{\beta g}^{4}-\frac{2}{15}\mu_{\beta g}\mu_{\alpha g}^{2} & 0 & -\frac{1}{15}\mu_{\beta g}^{2}\mu_{\alpha g}^{2} & 0 & 0\\
-\frac{1}{5}\mu_{\beta g}^{4}+\frac{1}{15}\mu_{\beta g}^{2}\mu_{\alpha g}^{2} & \frac{2}{15}\mu_{\beta g}^{2}\mu_{\alpha g}^{2} & \frac{2}{15}\mu_{\beta g}^{2}\mu_{\alpha g}^{2} & 0 & 0 & 0\\
-\frac{1}{15}\mu_{\beta g}^{4}+\frac{2}{15}\mu_{\beta g}^{2}\mu_{\alpha g}^{2} & \frac{4}{15}\mu_{\beta g}^{2}\mu_{\alpha g}^{2} & -\frac{1}{15}\mu_{\beta g}^{2}\mu_{\alpha g}^{2} & 0 & 0 & 0\\
0 & 0 & 0 & 0 & 0 & \frac{2}{15}\mu_{\beta g}^{2}\mu_{\alpha g}^{2}\\
0 & 0 & 0 & 0 & 0 & -\frac{1}{15}\mu_{\beta g}^{2}\mu_{\alpha g}^{2}\\
0 & 0 & 0 & 0 & \frac{2}{15}\mu_{\beta g}^{2}\mu_{\alpha g}^{2} & 0\\
0 & 0 & 0 & 0 & -\frac{1}{15}\mu_{\beta g}^{2}\mu_{\alpha g}^{2} & 0\end{array}\right]\left[\begin{array}{c}
\chi_{\beta\beta\beta\beta}(T)\\
\chi_{\alpha\alpha\beta\beta}(T)\\
\Re\{\chi_{\alpha\beta\alpha\beta}(T)\}\\
\Re\{\chi_{\beta\alpha\alpha\beta}(T)\}\\
\Im\{\chi_{\alpha\beta\alpha\beta}(T)\}\\
\Im\{\chi_{\beta\alpha\alpha\beta}(T)\}\end{array}\right] & = & \left[\begin{array}{c}
\Re\{\langle S_{\beta\beta}(T)\rangle_{zzzz}\}\\
\Re\{\langle S_{\beta\beta}(T)\rangle_{zzxx}\}\\
\Re\{\langle S_{\beta\alpha}(T)\rangle_{zzzz}\}\\
\Re\{\langle S_{\beta\alpha}(T)\rangle_{zzxx}\}\\
\Im\{\langle S_{\beta\beta}(T)\rangle_{zzzz}\}\\
\Im\{\langle S_{\beta\beta}(T)\rangle_{zzxx}\}\\
\Im\{\langle S_{\beta\alpha}(T)\rangle_{zzzz}\}\\
\Im\{\langle S_{\beta\alpha}(T)\rangle_{zzxx}\}\end{array}\right].\label{eq:rightside}\end{eqnarray}
}Inverting Eqs. (\ref{eq:leftside}) and (\ref{eq:rightside}) yields
most of the elements of $\chi(T)$ involving the single-exciton manifold.
Whereas the presented QPT for the homodimer is partial, no complicated
pulse shaping efforts need to be carried out. Instead, the requirement
is standard pulse polarization control achievable with current experimental
capabilities \cite{woutersen,perakis,zanni-njp,khalil:362,khaliljpc,moran-polarization,moran-polarization-ii,Schlau-Cohen27072010}. 

The transition dipole moments must be well characterized in order
to construct the matrices in Eqs. (\ref{eq:leftside}) and (\ref{eq:rightside}).
This requirement is self-consistently fulfilled by collecting the
spectra in the collinear and cross-polarized configurations. Notice
that $\chi_{\alpha\alpha\alpha\alpha}(T)$ and $\chi_{\beta\beta\alpha\alpha}(T)$
are exclusively monitored in the LHS, and $\chi_{\beta\beta\beta\beta}(T)$
and $\chi_{\alpha\alpha\beta\beta}(T)$ only detected in the RHS.
However, coherence transfer terms $\chi_{\alpha\beta\alpha\beta}(T)$
and $\chi_{\beta\alpha\alpha\beta}(T)$ are repeatedly monitored in
different peaks in both sides of the spectra. Due to this repetition,
there are redudant equations that allow for the self-consistent extraction
of the angle $\phi$ without compromising the inversion of the elements
of $\chi(T)$. Details about this parameter extraction are developed
in subsection E. For the time being, we assume that the information
about the transition dipoles is previously known.

\subsection{Stability of the Quantum Process Tomography protocol for a homodimer}

In order to characterize the stability of inversion of $\chi(T)$,
we can arrange Eqs. (\ref{eq:leftside}) and (\ref{eq:rightside})
into a single matrix equation $\boldsymbol{M}\vec{\chi}(T)=\vec{S}(T)$,
where $\boldsymbol{M}$ is a $16\times8$ matrix of dipole moments,
$\vec{\chi}(T)$ is a vector of 8 unknowns, and $\vec{S}(T)$ is a
vector of 16 real valued amplitudes extracted from the signal. Denoting
with $\parallel\cdot\parallel$ the spectral norm of a vector or a
matrix \cite{numerical}, we obtain a bound on the relative error
of the inverted vector $\vec{\chi}(T)$ which yields the QPT:

\begin{equation}
\frac{||\Delta\vec{\chi}(T)||}{||\vec{\chi}(T)||}\leq\kappa\frac{\parallel\Delta S(T)\parallel}{\parallel S(T)\parallel},\label{eq:bound}\end{equation}
where $\Delta\vec{\chi}(T)$ and $\Delta S(T)$ denote errors in $\vec{\chi}(T)$
and $S(T)$ upon inversion, respectively, and the \emph{condition
number} $\kappa$ is given by:

\begin{equation}
\kappa=\parallel\boldsymbol{M}\parallel\parallel\boldsymbol{M}^{-1}\parallel\label{eq:condition number}\end{equation}

The lowest possible value for a condition number is $\kappa=1$. In
Fig. (6), a plot of $\kappa$ vs. $\phi$ (red) is displayed. As expected,
very large values of $\kappa$, which denote unstable inversions,
are expected for systems where the site dipoles are aligned or antialigned,
when one of the eigenstates of $H_{S}$ becomes dark. We also indicate
a range of angles where $\kappa$ is below a reasonable threshold,
say $\kappa\leq15$ (blue horizontal line), which consists of angles
in the range $0.3\pi\leq\phi\leq0.7\pi$. Also, note that $\kappa$
is symmetric about the minimum of $\kappa$ at $\phi=\pi/2$, where
the best inversion is carried out with $\kappa\sim3.9$.

\begin{center}
\begin{figure}
\begin{centering}
\includegraphics[width=15cm]{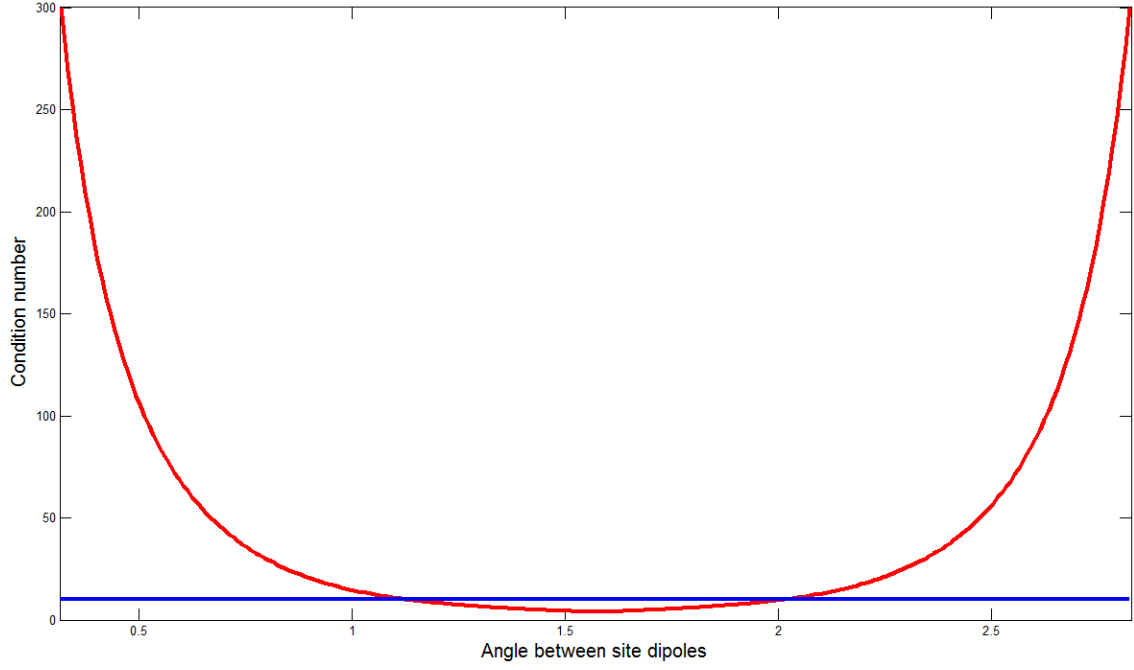} 
\par\end{centering}

\caption{Stability of the homodimer QPT protocol. As described in the text,
the QPT protocol depends on the inversion of a matrix which is a function
of transition dipole moments. The condition number of the matrix ($\kappa$)
vs. the angle between site dipoles ($\phi$) is plotted in red. The
blue line plots the constant value of $\kappa=15$, as a reference
to indicate that for the range of angles $0.3\pi\leq\phi\leq0.7\pi$,
the condition number is below that value.}

\end{figure}

\par\end{center}

\subsection{Analytical expressions for $\chi(T)$}

To gain insights into the QPT protocol, we shall derive explicit expressions
for the elements of $\chi(T)$ in terms of the amplitudes of the spectra
and the angle $\phi$ between the dipoles. First, we substitute Eqs.
(\ref{eq:dipole equals d}) and (\ref{eq:dipoles homodimer-1}) into
the LHS of the spectra through Eq. (\ref{eq:leftside}). The following
expressions are obtained after inverting the resulting matrix equation:

\begin{eqnarray}
\chi_{gg\alpha\alpha}(T)-1 & = & -\frac{\sec^{2}(\phi/2)}{(iC^{3})(20d^{4})}\nonumber \\
 &  & \times\Re\{\langle S_{\alpha\alpha}(T)\rangle_{zzzz}+2\langle S_{\alpha\alpha}(T)\rangle_{zzxx}\nonumber \\
 &  & +\langle S_{\alpha\beta}(T)\rangle_{zzzz}+2\langle S_{\alpha\beta}(T)\rangle_{zzxx}\}\label{eq:xggaa}\end{eqnarray}
\\
\begin{eqnarray}
\chi_{\alpha\alpha\alpha\alpha}(T) & = & -\frac{\sec^{2}(\phi/2)}{(iC^{3})(20d^{4})}\nonumber \\
 &  & \times\Re\{(\cos\phi-1)(\langle S_{\alpha\alpha}(T)\rangle_{zzzz}+2\langle S_{\alpha\alpha}(T)\rangle_{zzxx}\nonumber \\
 &  & +\cos\phi(\langle S_{\alpha\beta}(T)\rangle_{zzzz}+2\langle S_{\alpha\beta}(T)\rangle_{zzxx})\}\label{eq:xaaaa}\end{eqnarray}

`\begin{eqnarray}
\chi_{\beta\beta\alpha\alpha}(T) & = & \frac{\sec^{2}(\phi/2)}{(iC^{3})(20d^{4})}\nonumber \\
 &  & \times\Re\{\cos\phi(\langle S_{\alpha\alpha}(T)\rangle_{zzzz}+2\langle S_{\alpha\alpha}(T)\rangle_{zzxx})\nonumber \\
 &  & +(\cos\phi+1)(\langle S_{\alpha\beta}(T)\rangle_{zzzz}+2\langle S_{\alpha\beta}(T)\rangle_{zzxx})\}\label{eq:xbbaa}\end{eqnarray}

\begin{eqnarray}
\Re\{\chi_{\alpha\beta\alpha\beta}(T)\} & = & \frac{1}{(iC^{3})(20d^{4})}\nonumber \\
 &  & \times\Re\{2(\langle S_{\alpha\alpha}(T)\rangle_{zzzz}+2\langle S_{\alpha\alpha}(T)\rangle_{zzxx})\nonumber \\
 &  & +(\cot^{2}\frac{\phi}{2}+2\tan^{2}\frac{\phi}{2}+5)\langle S_{\alpha\beta}(T)\rangle_{zzzz}\nonumber \\
 &  & -(3\cot^{2}\frac{\phi}{2}+\tan^{2}\frac{\phi}{2})\langle S_{\alpha\beta}(T)\rangle_{zzxx}\}\label{eq:xabab}\end{eqnarray}

\begin{eqnarray}
\Re\{\chi_{\beta\alpha\alpha\beta}(T)\} & = & \frac{1}{(iC^{3})(20d^{4})}\nonumber \\
 &  & \times\Re\{(\cot^{2}\frac{\phi}{2}+2\tan^{2}\frac{\phi}{2}+1)\langle S_{\alpha\alpha}(T)\rangle_{zzzz}\nonumber \\
 &  & -(3\cot^{2}\frac{\phi}{2}+\tan^{2}\frac{\phi}{2}+8)\langle S_{\alpha\alpha}(T)\rangle_{zzxx}\nonumber \\
 &  & -2(\langle S_{\alpha\beta}(T)\rangle_{zzzz}+2\langle S_{\alpha\beta}(T)\rangle_{zzxx})\}\label{eq:xbaab}\end{eqnarray}

\begin{eqnarray}
\Im\{\chi_{\alpha\beta\alpha\beta}(T)\} & = & \frac{15}{(iC^{3})(8d^{4})}\Im\{(\csc^{2}\frac{\phi}{2}\sec^{2}\frac{\phi}{2})\langle S_{\alpha\beta}(T)\rangle_{zzzz}\}\nonumber \\
 & = & -\frac{15}{(iC^{3})(4d^{4})}\Im\{(\csc^{2}\frac{\phi}{2}\sec^{2}\frac{\phi}{2})\langle S_{\alpha\beta}(T)\rangle_{zzxx}\}\label{eq:xabab-im}\end{eqnarray}

\begin{eqnarray}
\Im\{\chi_{\beta\alpha\alpha\beta}(T)\} & = & \frac{15}{(iC^{3})(8d^{4})}\Im\{(\csc^{2}\frac{\phi}{2}\sec^{2}\frac{\phi}{2})\langle S_{\alpha\alpha}(T)\rangle_{zzzz}\}\nonumber \\
 & = & -\frac{15}{(iC^{3})(4d^{4})}\Im\{(\csc^{2}\frac{\phi}{2}\sec^{2}\frac{\phi}{2})\langle S_{\alpha\alpha}(T)\rangle_{zzxx}\}\label{eq:xbaab-im}\end{eqnarray}

Similarly, the RHS of the spectra through Eq. (\ref{eq:rightside})
yield:

\begin{eqnarray}
\chi_{gg\beta\beta}(T)-1 & = & -\frac{\csc^{2}(\phi/2)}{(iC^{3})(20d^{4})}\nonumber \\
 &  & \times\Re\{\langle S_{\beta\beta}(T)\rangle_{zzzz}+2\langle S_{\beta\beta}(T)\rangle_{zzxx}\nonumber \\
 &  & +\langle S_{\beta\alpha}(T)\rangle_{zzzz}+2\langle S_{\beta\alpha}(T)\rangle_{zzxx}\}\label{eq:xggbb}\end{eqnarray}
\\
\begin{eqnarray}
\chi_{\beta\beta\beta\beta}(T) & = & \frac{\csc^{2}(\phi/2)}{(iC^{3})(20d^{4})}\nonumber \\
 &  & \times\Re\{(\cos\phi+1)(\langle S_{\beta\beta}(T)\rangle_{zzzz}+2\langle S_{\beta\beta}(T)\rangle_{zzxx}\nonumber \\
 &  & +\cos\phi(\langle S_{\beta\alpha}(T)\rangle_{zzzz}+2\langle S_{\beta\alpha}(T)\rangle_{zzxx})\}\label{eq:xbbbb}\end{eqnarray}

`\begin{eqnarray}
\chi_{\alpha\alpha\beta\beta}(T) & =- & \frac{\csc^{2}(\phi/2)}{(iC^{3})(20d^{4})}\nonumber \\
 &  & \times\Re\{\cos\phi(\langle S_{\beta\beta}(T)\rangle_{zzzz}+2\langle S_{\beta\beta}(T)\rangle_{zzxx})\nonumber \\
 &  & +(\cos\phi-1)(\langle S_{\beta\alpha}(T)\rangle_{zzzz}+2\langle S_{\beta\alpha}(T)\rangle_{zzxx})\}\label{eq:xaabb}\end{eqnarray}

\begin{eqnarray}
\Re\{\chi_{\alpha\beta\alpha\beta}(T)\} & = & \frac{1}{(iC^{3})(20d^{4})}\nonumber \\
 &  & \times\Re\{2(\langle S_{\beta\beta}(T)\rangle_{zzzz}+2\langle S_{\beta\beta}(T)\rangle_{zzxx})\nonumber \\
 &  & +(\tan^{2}\frac{\phi}{2}+2\cot^{2}\frac{\phi}{2}+5)\langle S_{\beta\alpha}(T)\rangle_{zzzz}\nonumber \\
 &  & -(3\tan^{2}\frac{\phi}{2}+\cot^{2}\frac{\phi}{2})\langle S_{\beta\alpha}(T)\rangle_{zzxx}\}\label{eq:xabab-1}\end{eqnarray}

\begin{eqnarray}
\Re\{\chi_{\beta\alpha\alpha\beta}(T)\} & = & \frac{1}{(iC^{3})(20d^{4})}\nonumber \\
 &  & \times\Re\{(\tan^{2}\frac{\phi}{2}+2\cot^{2}\frac{\phi}{2}+1)\langle S_{\beta\beta}(T)\rangle_{zzzz}\nonumber \\
 &  & -(3\tan^{2}\frac{\phi}{2}+\cot^{2}\frac{\phi}{2}+8)\langle S_{\beta\beta}(T)\rangle_{zzxx}\nonumber \\
 &  & -2(\langle S_{\beta\alpha}(T)\rangle_{zzzz}+2\langle S_{\beta\alpha}(T)\rangle_{zzxx})\}\label{eq:xbaab-1}\end{eqnarray}

\begin{eqnarray}
\Im\{\chi_{\alpha\beta\alpha\beta}(T)\} & = & -\frac{15}{(iC^{3})(8d^{4})}\Im\{(\sec^{2}\frac{\phi}{2}\csc^{2}\frac{\phi}{2})\langle S_{\beta\alpha}(T)\rangle_{zzzz}\}\nonumber \\
 & = & \frac{15}{(iC^{3})(4d^{4})}\Im\{(\sec^{2}\frac{\phi}{2}\csc^{2}\frac{\phi}{2})\langle S_{\beta\alpha}(T)\rangle_{zzxx}\}\label{eq:im-xabab-1}\end{eqnarray}

\begin{eqnarray}
\Im\{\chi_{\beta\alpha\alpha\beta}(T)\} & = & -\frac{15}{(iC^{3})(8d^{4})}\Im\{(\sec^{2}\frac{\phi}{2}\csc^{2}\frac{\phi}{2})\langle S_{\beta\beta}(T)\rangle_{zzzz}\}\nonumber \\
 & = & \frac{15}{(iC^{3})(4d^{4})}\Im\{(\sec^{2}\frac{\phi}{2}\csc^{2}\frac{\phi}{2})\langle S_{\beta\beta}(T)\rangle_{zzxx}\}\label{eq:im-xbaab-1}\end{eqnarray}
As expected, the RHS equations above can be obtained from the LHS
equations upon the substitutions $\alpha\to\beta$, $\beta\to\alpha$,
and $\phi\to\pi-\phi$ (see Eqs. (\ref{eq:dipoles homodimer-1})). 

From Eqs. (\ref{eq:xabab-im}) and (\ref{eq:im-xabab-1}), we notice
that the imaginary parts of $\langle S_{\alpha\beta}(T)\rangle_{zzzz},\langle S_{\alpha\beta}(T)\rangle_{zzxx},\langle S_{\beta\alpha}(T)\rangle_{zzxx},\langle S_{\beta\alpha}(T)\rangle_{zzxx}$
are all exclusively proportional to $\Im\{\chi_{\alpha\beta\alpha\beta}(T)\}$.
The least-squares solution for $\Im\{\chi_{\alpha\beta\alpha\beta}(T)\}$
using Eqs. (\ref{eq:leftside}) and (\ref{eq:rightside}) is simply
the average of these four values; the analogous conclusion holds for
$\Im\{\chi_{\beta\alpha\alpha\beta}(T)\}$. The real parts, $\Re\{\chi_{\alpha\beta\alpha\beta}(T)\},\Re\{\chi_{\beta\alpha\alpha\beta}(T)\}$
satisfy less trivial relationships. Each of them appears twice in
the equations, once in the RHS and another time in the LHS of the
spectra. It is this redundancy in the spectral information what allows
for the extraction of the angle between the dipoles.

\subsection{Determination of the angle $\phi$ between the two dipoles}

Equating the values of $\Re\{\chi_{\alpha\beta\alpha\beta}(T)\}$
from the expressions in Eqs. (\ref{eq:xabab}) and (\ref{eq:xabab-1})
yield the following quadratic equation in $\xi=\tan^{2}\left(\frac{\phi}{2}\right)$:

\begin{eqnarray}
 &  & \Re\{[\langle S_{\beta\alpha}(T)\rangle_{zzzz}-3\langle S_{\beta\alpha}(T)\rangle_{zzxx}-2\langle S_{\alpha\beta}(T)\rangle_{zzzz}+\langle S_{\alpha\beta}(T)\rangle_{zzxx}]\xi^{2}\nonumber \\
 &  & +[5\langle S_{\beta\alpha}(T)\rangle_{zzzz}+2\langle S_{\beta\beta}(T)\rangle_{zzzz}+4\langle S_{\beta\beta}(T)\rangle_{zzxx}\nonumber \\
 &  & -5\langle S_{\alpha\beta}(T)\rangle_{zzzz}-2\langle S_{\alpha\alpha}(T)\rangle_{zzzz}-4\langle S_{\alpha\alpha}(T)\rangle_{zzxx}]\xi\nonumber \\
 &  & +[2\langle S_{\beta\alpha}(T)\rangle_{zzzz}-\langle S_{\beta\alpha}(T)\rangle_{zzxx}-\langle S_{\alpha\beta}(T)\rangle_{zzzz}+3\langle S_{\alpha\beta}(T)\rangle_{zzxx}]\}\nonumber \\
 & = & 0.\label{eq:quadratic}\end{eqnarray}

A similar expression can be found for $\Re\{\chi_{\beta\alpha\alpha\beta}(T)\}$
from Eqs. (\ref{eq:xbaab}) and (\ref{eq:xbaab-1}):

\begin{eqnarray}
 &  & \Re\{[2\langle S_{\alpha\alpha}(T)\rangle_{zzzz}-\langle S_{\alpha\alpha}(T)\rangle_{zzxx}-\langle S_{\beta\beta}(T)\rangle_{zzzz}+3\langle S_{\beta\beta}(T)\rangle_{zzxx}]\xi^{2}\nonumber \\
 &  & +[\langle S_{\alpha\alpha}(T)\rangle_{zzzz}-8\langle S_{\alpha\alpha}(T)\rangle_{zzxx}-2\langle S_{\alpha\beta}(T)\rangle_{zzzz}-4\langle S_{\alpha\beta}(T)\rangle_{zzxx}\nonumber \\
 &  & -\langle S_{\beta\beta}(T)\rangle_{zzzz}+8\langle S_{\beta\beta}(T)\rangle_{zzxx}+2\langle S_{\beta\alpha}(T)\rangle_{zzzz}+4\langle S_{\beta\alpha}(T)\rangle_{zzxx}]\xi\nonumber \\
 &  & +[\langle S_{\alpha\alpha}(T)\rangle_{zzzz}-3\langle S_{\alpha\alpha}(T)\rangle_{zzxx}-2\langle S_{\beta\beta}(T)\rangle_{zzzz}+\langle S_{\beta\beta}(T)\rangle_{zzxx}]\}\nonumber \\
 & = & 0.\label{eq:quadratic-1}\end{eqnarray}

The identities in Eqs. (\ref{eq:quadratic}) and (\ref{eq:quadratic-1})
are remarkable in the sense that they are satisfied at \emph{every}
waiting time $T$: they do not depend on short time coherent dynamics.
By monitoring the peak amplitudes from the spectra arising from the
two different polarization configurations, the angle $\phi$ between
the two site dipoles can be readily extracted using either expression.
This determination is robust because it can be repeated for every
value of $T$ for which the signal has been collected.

\subsection{Summary of Quantum Process Tomography protocol for a coupled homodimer}

We proceed to summarize the algorithm of the QPT protocol for a coupled
homodimer.
\begin{enumerate}
\item Obtain the amplitudes $\langle S_{mn}(T)\rangle_{e_{1}e_{2}e_{3}e_{4}}$
for $m,n\in\{\alpha,\beta\}$ and the two polarization settings $(\boldsymbol{e}_{1},\boldsymbol{e}_{2},\boldsymbol{e}_{3},\boldsymbol{e}_{4})=(\boldsymbol{z},\boldsymbol{z},\boldsymbol{z},\boldsymbol{z}),(\boldsymbol{x},\boldsymbol{x},\boldsymbol{z},\boldsymbol{z})$.
This information can be extracted from the two respective polarization
controlled 2D-ES, $\langle S(\omega_{\tau},T,\omega_{t})\rangle_{e_{1}e_{2}e_{3}e_{4}}$.
For simplicity, all the pulses are taken to be of the same duration
(short compared to the timescales of excited state dynamics). 
\item Extract the angle $\phi$ from the data from step 1 and Eqs. (\ref{eq:quadratic})
and (\ref{eq:quadratic-1}).
\item Plug in information obtained from step 1 and the angle $\phi$ from
step 2 into the expressions for the elements of $\chi(T)$ in Eqs.
(\ref{eq:xggaa})-(\ref{eq:im-xbaab-1}). Some important observations:
(a) These expressions are all proportional to the factor $(C^{3}d^{4})^{-1}$.
The norm of the dipole $d$ can be extracted from the intensity of
the absorption spectrum of the monomer. If this information is not
readily available, the results are known up to this constant factor.
(b) By construction from Eq. (\ref{eq:quadratic}), the calculated
value of $\Re\{\chi_{\alpha\beta\alpha\beta}(T)\}$ will be the same
using either Eqs. (\ref{eq:xabab}) or (\ref{eq:xabab-1}). The same
holds for Eq. (\ref{eq:quadratic-1}), $\Re\{\chi_{\beta\alpha\alpha\beta}(T)\}$
and Eqs. (\ref{eq:xabab}) and (\ref{eq:xabab-1}).
\end{enumerate}

\subsection{Numerical example}

In this subsection, we illustrate the described QPT protocol with
a model homodimer. Marcus and coworkers have recently reported a synthetic
system of porphyrin molecules which self-assemble into homodimers
under the presence of liposomes \cite{marcus}. The parameters of
this system, extracted from phase-modulation electronic coherence
spectroscopy (PM-ECS), are $\bar{\omega}=16633\, cm^{-1},$ $J=175\, cm^{-1}$,
and $\phi=65^{o}$. The transition energies to the eigenstates are
$\omega_{\alpha g}=16458\, cm^{-1}$ and $\omega_{\beta g}=16808\, cm^{-1}$.
Information on the spectral density of this system is not available
in the literature yet. We adopt a simple system-bath model based on
the secular Redfield approach and independent bath approximation (IBA)
for each site. The weak system-bath model is reasonable since porphyrins
are rigid molecules which change their structures minimally upon electronic
excitation. The IBA must be reexamined, since the liposome media guarantee
a bosonic bath that could be strongly correlated in both sites. Nevertheless,
the purpose of this example is not to provide an exact account of
the excited state dynamics of this system, but rather an illustration
of the QPT protocol using reasonable timescales that one might encounter
in a realistic setting. A careful study of the precise bath-induced
dynamics in this system is beyond the scope of this study and shall
be addressed in future work in collaboration with an experimental
realization.

We consider a harmonic bath with an Ohmic spectral density: $J(\omega)=\frac{\lambda}{\omega_{c}}\omega e^{-\omega/\omega_{c}}$,
with $\lambda=100\, cm^{-1}$, $\omega_{c}=150\, cm^{-1}$ at a temperature
$T=273\, K$. Identical baths are assumed to be diagonally and linearly
coupled to each of the sites. We closely follow the calculation reported
in \cite{nonperturbativepisliakov} and adapted in \cite{yuenzhou}.
The dynamics of the total excitonic system, which is a proper density
matrix, is governed by the following equation of motion:

\begin{equation}
\dot{\rho}(T)=-i[H_{S},\rho(T)]-\mathcal{R}\rho(T)\label{eq:redfield}\end{equation}
where $\mathcal{R}$ denotes the time-independent sparse dissipative
superoperator containing only a few non-zero elements listed in Table
3. Since $\dot{\rho}(T)$ only depends on $\rho(T)$ and not on the
value of the quantum state at previous times, the simulated dynamics
are Markovian.\\
~

\begin{center}
\begin{tabular}{cc}
\hline 
\multicolumn{2}{c}{TABLE 3. Values (in $fs^{-1}$) of non-zero rates of the secular Redfield
tensor }\tabularnewline
\hline
$R_{\beta\beta\alpha\alpha}$ & $8.02\times10^{-4}$\tabularnewline
$R_{\alpha\alpha\beta\beta}=e^{-\omega_{\alpha\beta}/k_{B}T}R_{\beta\beta\alpha\alpha}$ & $5.07\times10^{-3}$\tabularnewline
$R_{\alpha\beta\alpha\beta}=R_{\beta\alpha\beta\alpha}$ & $2.93\times10^{-3}$\tabularnewline
$R_{\alpha g\alpha g}=R_{g\alpha g\alpha}=R_{f\alpha f\alpha}=R_{\alpha f\alpha f}$ & $1.23\times10^{-2}$\tabularnewline
$R_{\beta g\beta g}=R_{g\beta g\beta}=R_{f\beta f\beta}=R_{\beta f\beta f}$ & $1.45\times10^{-2}$\tabularnewline
$R_{fgfg}=R_{gfgf}$ & $4.77\times10^{-2}$\tabularnewline
\hline
\end{tabular}\\
~
\par\end{center}

It is well known that the secular Redfield equations guarantee thermal
equilibrium since the population transfer rates satisfy $R_{\alpha\alpha\beta\beta}/R_{\beta\beta\alpha\alpha}=e^{-\omega_{\alpha\beta}/k_{B}T}$,
where $k_{B}$ is the Boltzmann constant. Also, $R_{fgfg}$ will not
be relevant for the calculations, as coherences between the ground
state and the biexciton are never created in the PE experiment. The
free-induction decay rates for the coherence and echo intervals will
be taken for simplicity to be the same, $\Gamma_{\alpha g}\approx\Gamma_{\beta g}\approx\frac{1}{2}(R_{\alpha g\alpha g}+R_{\beta g\beta g})$.
This restriction is by no means necessary, but will simplify the simulations
below. 

The non-zero elements of $\chi(T)$ for the single-exciton manifold
are presented in Table 4.

~

\begin{center}
\begin{tabular}{cc}
\hline 
\multicolumn{2}{c}{TABLE 4. Nonzero elements of $\chi(T)$ involving single-exciton states
for the secular Redfield model}\tabularnewline
\hline
$\chi_{\alpha\alpha\alpha\alpha}(T)$ & $1-e^{-R_{\beta\beta\alpha\alpha}T}$\tabularnewline
$\chi_{\beta\beta\alpha\alpha}(T)$ & $e^{-R_{\beta\beta\alpha\alpha}T}$\tabularnewline
$\chi_{\alpha\alpha\beta\beta}(T)$ & $e^{-R_{\alpha\alpha\beta\beta}T}$\tabularnewline
$\chi_{\beta\beta\alpha\alpha}(T)$ & $1-e^{-R_{\beta\beta\alpha\alpha}T}$\tabularnewline
$\chi_{\alpha\beta\alpha\beta}(T)=(\chi_{\beta\alpha\beta\alpha}(T))^{*}$ & $e^{-i\omega_{\alpha\beta}T}e^{-R_{\alpha\beta\alpha\beta}T}$\tabularnewline
\hline
\end{tabular}
\par\end{center}

~

In this particular calculation, coupling to the photon bath has been
ignored beyond the ultrashort pulses, as spontaneous emission occurs
in the order of nanoseconds, i.e. $\chi_{gg\alpha\alpha}(T)=\chi_{gg\beta\beta}(T)=0$.
Therefore, Eq. (\ref{eq:trace}) can be readily confirmed as $\chi_{\alpha\alpha\alpha\alpha}(T)+\chi_{\beta\beta\alpha\alpha}(T)=\chi_{\alpha\alpha\beta\beta}(T)+\chi_{\beta\beta\alpha\alpha}(T)=1$.
Also, due to the secular approximation, $\chi_{\beta\alpha\alpha\beta}(T)=(\chi_{\alpha\beta\beta\alpha}(T))^{*}=0$.
The picture of the secular Redfield equations is very simple and provides
transparent means for understanding the QPT protocol for the homodimer:
The evolution of populations and the coherences independently satisfy
standard first-order kinetic equations, leading to multiexponential
integrated dynamics.

In Fig.7, we display the calculated 2D-ES of this model system. We
consider the three pulses to be identical, centered about $\omega_{1}=\omega_{2}=\omega_{3}=16546\, cm^{-1}$,
of $FWHM=20\, fs$, i.e., $\sigma=8.49\, fs$, which amount to an
equal excitation amplitude $C$ for both $|\alpha\rangle$ and $|\beta\rangle$.
Rabi oscillations for a coherent superposition between $|\alpha\rangle$
and $|\beta\rangle$ occur with a period $T_{c}=47.5\, fs$. We present
several snapshots of the real and imaginary parts of the spectra at
values of waiting time $T$ corresponding to multiples of $T_{c}/2$,
skipping $T_{c}=0$, as our theory has avoided pulse overlap effects.
In principle, as Eqs. (\ref{eq:sab}) and (\ref{eq:sba}) indicate,
the excitonic quantum beats associated with the term $\chi_{\alpha\beta\alpha\beta}(T)$
can be monitored by looking at either cross peak of the spectra. This
feature is subtly manifested in every column of the figure, but more
easily perceived in the real part the $zzxx$ spectrum, where the
peak at $(\omega_{\beta},\omega_{\alpha})$ changes from red to yellow/green
every interval $T_{c}/2$ before the bosonic bath has washed out significant
portion of the coherent dynamics at about $T=5T_{c}$. Also, incoherent
population transfer primarily from $|\beta\rangle$ to $|\alpha\rangle$
(downhill) manifests as a decrease in amplitude of the peak at $(\omega_{\beta},\omega_{\beta})$
due to ESA and an increase in $(\omega_{\beta},\omega_{\alpha})$
due to SE \cite{minhaengbook}. This effect can also be more obviously
seen in the real part of $zzxx$ spectrum.

\begin{center}
\begin{figure}
\begin{centering}
\includegraphics[width=15cm]{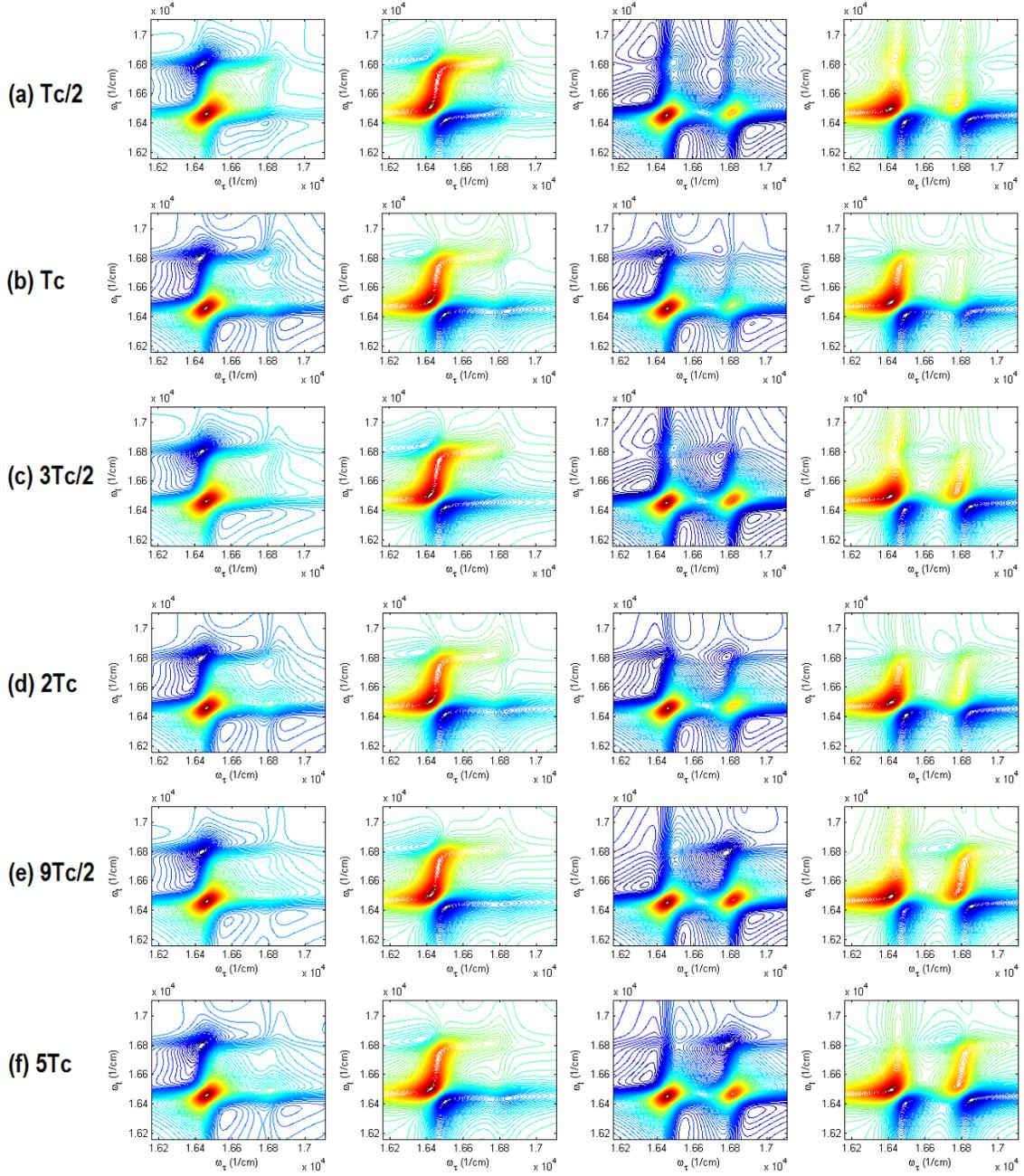} 
\par\end{centering}

\caption{2D-ES for coupled porphyrin homodimer with secular Redfield model.
From left to right, we show the real and imaginary parts of the spectrum
with $zzzz$ polarization setting (first and second columns), and
with $zzxx$ setting (third and fourth columns). Each row represents
a particular waiting time $T$, corresponding to (a) $T_{c}/2$, (b)
$T_{c}$, (c) $3T_{c}/2$, (d) $2T_{c}$, (e) $9T_{c}/2$, (f) $5T_{c}$,
where $T_{c}=47.5\, fs$ is the period for one Rabi oscillation between
$|\alpha\rangle$ and $|\beta\rangle$. The colormap is such that
red is associated with positive numbers, green with values about zero,
and blue with negative numbers.}

\end{figure}

\par\end{center}

From the simulated spectra $\langle S(\omega_{\tau},T,\omega_{t})\rangle_{e_{1}e_{2}e_{3}e_{4}}$,
the extraction of the terms $\langle S_{mn}(T)\rangle_{e_{1}e_{2}e_{3}e_{4}}$
is achieved with high fidelity (>99\%) by a nonlinear optimization
routine based on the simplex search method with bound constraints
\cite{siam}. The signals are fitted to a sum of four different resonances
as in the isotropically averaged version of Eq. (\ref{eq:2dspectrum expression}).
The parameters $\omega_{mg}$, $\omega_{ng}$, $\Gamma$, and $\langle S_{mn}(T)\rangle_{e_{1}e_{2}e_{3}e_{4}}$
are reconstructed from 2D-ES with a grid spacing of $\Delta\omega_{\tau}=\Delta\omega_{t}=1\, cm^{-1}$
and a grid size of $1050\, cm^{-1}$ for every axis. We present the
results of this calculation in Figure 8. Notice that the imaginary
parts of the diagonal peaks are zero since no terms of the form $\chi_{\beta\alpha\alpha\beta}(T)=(\chi_{\alpha\beta\beta\alpha}(T))^{*}$are
considered in the secular Redfield theory, and population transfer
terms are purely real.

\begin{center}
\begin{figure}
\begin{centering}
\includegraphics[width=15cm]{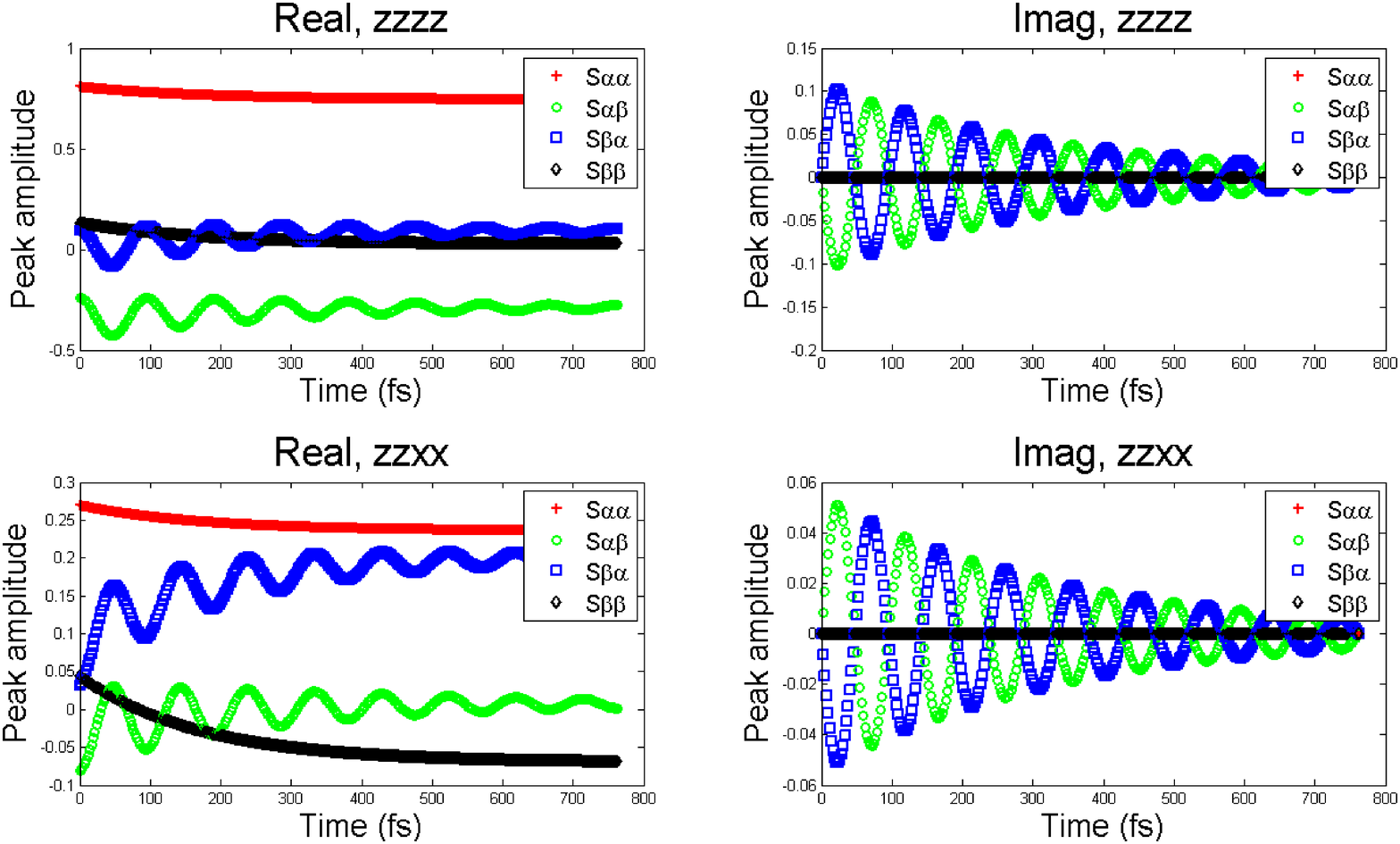} 
\par\end{centering}

\caption{Traces of $\langle S_{mn}(T)\rangle_{e_{1}e_{2}e_{3}e_{4}}$ for $m,n\in\{\alpha,\beta\}$
and $(\boldsymbol{e}_{1},\boldsymbol{e}_{2},\boldsymbol{e}_{3},\boldsymbol{e}_{4})\in\{(\boldsymbol{z},\boldsymbol{z},\boldsymbol{z},\boldsymbol{z}),(\boldsymbol{z},\boldsymbol{z},\boldsymbol{x},\boldsymbol{x})\}$.
(a) Real and (b) imaginary parts for the $zzzz$ configuration; (c)
real and (d) imaginary parts for the $zzxx$ configuration. Each of
the plots shows the evolution of the peak amplitudes $S_{\alpha\alpha}(T)$
(red crosses), $S_{\alpha\beta}(T)$ (green circles), $S_{\beta\alpha}(T)$
(blue squares), and $S_{\beta\beta}(T)$ (black diamonds).}

\end{figure}

\par\end{center}

Eq. (\ref{eq:quadratic}) is solved at every waiting time $T\in[T_{c}/2,10T_{c}]$
yielding the roots $\xi=1.000,0.4059$ for every $T$, without variance
after the fourth decimal digit, indicating its robustness for the
inversion of $\phi$. The same exercise with Eq. (\ref{eq:quadratic-1})
gives $\xi=-0.4059,0.4059$. These values of $\xi$ imply $\phi=90^{0},65^{0},65^{0},86.3i^{0}$,
respectively. The last value can be discarded for it is not even a
real number. The value of $\phi$ must be a root of both equations.
Therefore, we can also discard $90^{0}$, since it is only solution
of the first equation, but not of the second. The result $\phi=65^{0}$
follows unambiguously, as expected. 

Finally, the terms $\langle S_{mn}(T)\rangle_{e_{1}e_{2}e_{3}e_{4}}$
and the angle $\phi$ allow for the evaluation of the elements of
$\chi(T)$ which are extractable for the homodimer. Figure 9 shows
that this reconstruction coincides with the analytical expressions
presented in Table 4. The population decay terms $\chi_{\alpha\alpha\alpha\alpha}(T),\chi_{\beta\beta\beta\beta}(T)$
both start at 1 and reach 0 exponentially, the second faster than
the first, since $|\beta\rangle$ is the excitonic state of higher
energy. The population transfer terms $\chi_{\alpha\alpha\beta\beta}(T),\chi_{\beta\beta\alpha\alpha}(T)$
are complementary to the former ones, with the transfer from $|\beta\rangle$
to $|\alpha\rangle$ being faster for the same reasons just mentioned.
The coherence term decays exponentially, with real and imaginary parts
$\pi/2$ phase shifted one from another. The calculated timescale
of this decay (hundreds of femtoseconds) is similar to the one inferred
from the experiment reported by Lee and coworkers, where a superposition
of excitons in the bacteriopheophytin and bacteriochlorophyll sites
in the reaction center of purple bacteria is monitored indirectly
through a two-color experiment \cite{HohjaiLee06082007}. 

\begin{center}
\begin{figure}
\begin{centering}
\includegraphics[width=15cm]{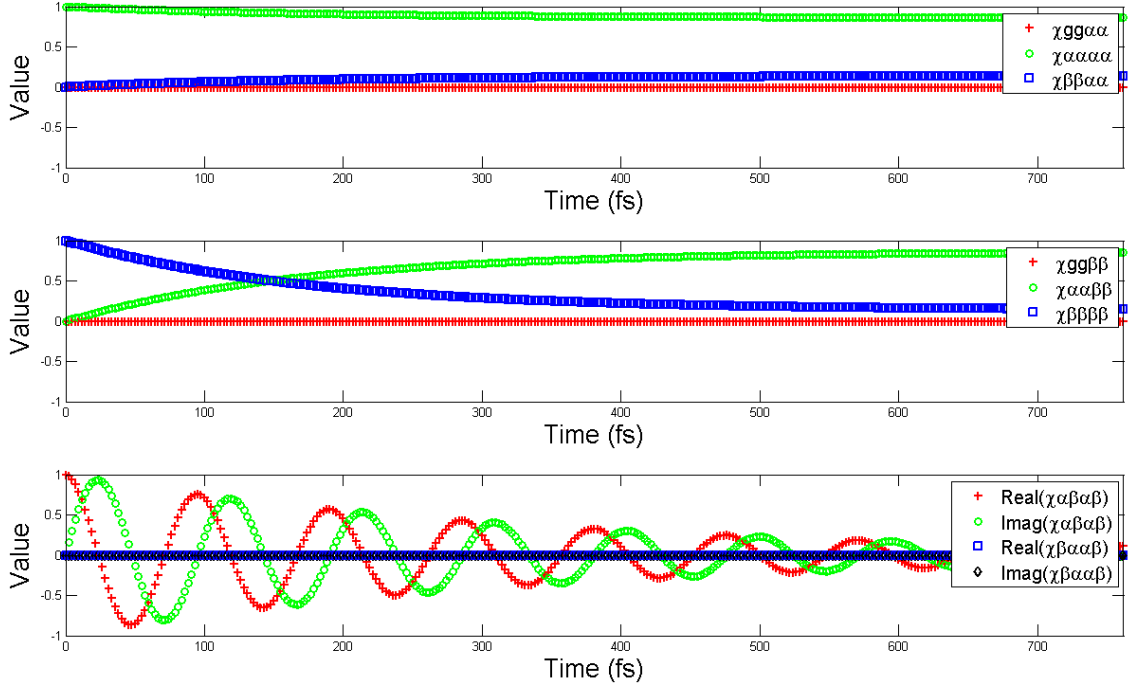} 
\par\end{centering}

\caption{Extractable elements of $\chi(T)$ for homodimer. (a) Processes starting
from $|\alpha\rangle\langle\alpha|$: $\chi_{gg\alpha\alpha}(T)$
(red crosses), $\chi_{\alpha\alpha\alpha\alpha}(T)$ (green circles),
$\chi_{\beta\beta\alpha\alpha}(T)$ (blue squares). (b) Processes
starting from $|\beta\rangle\langle\beta|$: $\chi_{gg\beta\beta}(T)$
(red crosses), $\chi_{\alpha\alpha\beta\beta}(T)$ (green circles),
$\chi_{\beta\beta\beta\beta}$ (blue squares). (c) Processes indicating
coherence transfer: $\Re\{\chi_{\alpha\beta\alpha\beta}(T)\}$ (red
crosses), $\Im\{\chi_{\alpha\beta\alpha\beta}(T)\}$ (green circles),
$\Re\{\chi_{\beta\alpha\alpha\beta}(T)\}$ (blue squares), $\Im\{\chi_{\beta\alpha\alpha\beta}(T)\}$
(black diamonds). }

\end{figure}

\par\end{center}

\section{Discussion}

In the present article, we have outlined a general theory for carrying
out a QPT for a molecular dimer using the information contained in
various frequency and polarization controlled 2D-ES. We started by
providing the basic concepts of QPT, and operationally defined a QPT
as a protocol to extract the process matrix $\chi(T)$, which in principle
completely characterizes a quantum black box, in our case, the box
being the single-exciton manifold of the dimer. After reviewing the
model Hamiltonian as well as the transition dipole moments of an excitonic
dimer, we adapted the QPT theory presented in our previous work, where
the nonlinear polarization was analyzed in real time for single values
of $\tau$ and $t$ times (see Eqs. (\ref{eq:total polarization}),
(\ref{eq:with alpha}), and (\ref{eq:with beta})) \cite{yuenzhou},
to the more standard and visual Fourier transformed signal collected
along several values of these interval times. The central result of
this exercise was Eqs. (\ref{eq:saa}), (\ref{eq:sab}), (\ref{eq:sbb}),
and (\ref{eq:sba}), which from a purist stanpoint completes the QPT
effort: The peaks in a heterodyne-detected 2D-ES can be expressed
as linear combinations of elements of the process matrix $\chi(T)$.
This information can be distilled by carrying out several experiments
alternating the frequency components of the pulses as well as their
polarization. By setting up a system of linear equations with this
data, a linear algebraic routine yields the inversion of $\chi(T)$
for every waiting time $T$. In order to get a more intuitive picture
of this procedure, the particular case of a homodimer was studied
in detail. The degeneracies of this system yield a perpendicular set
of transition dipole moments which considerably simplify the theory
(see Fig. 5). It was shown that under isotropic average of the signal,
no population to coherence processes or viceversa can be monitored,
impeding a full QPT for the single-exciton manifold of this system.
However, the partially achievable QPT is very simple, robust with
respect to transition dipole moment parameters as long as they are
not aligned or antialigned, and readily implemented without pulse
shaping. The only requirement is the collection of two polarization
controlled 2D-ES. Numerical examples with a model homodimer validated
the presented theory.

The possibilities that QPT opens for the study of excited state dynamics
in condensed molecular systems are as vast as the information acquired
at the amplitude level of the evolving quantum state of the probed
system. On the one hand, with the peaks in the 2D spectrum indicating
a plethora of pathways in Liouville space, understanding of the dynamics
is undoubtedly enhanced by the dissection of these peaks into processes
described by the $\chi(T)$ matrix. On the other hand, a plethora
of questions can be addressed with this information, for instance:
Is a Markovian description accurate? \cite{AkihitoIshizaki10132009,ishizakijcp,ChengEngel,renger:9997}
If not, what is the degree of non-Markovianity of the dynamics? \cite{breuernonmarkovian,patricknonmarkovian}
If it is Markovian, is the secular Redfield equation appropriate or
are non-secular processes important? \cite{ungar}. Is there any degree
of entanglement in the quantum states produced in the single exciton
manifold upon photoexcitation? \cite{sarovar,ishizakinjp} What is
the rate of decoherence of a quantum superposition between excitonic
states? \cite{mancalspec,olsinamancal,abramaviciuslindblad,hayesengel,schlaujpcb} 

A few aspects have not been fully addressed with respect to the implementation
of QPT of a molecular dimer. These issues will be carefully studied
in future publications in collaboration with experimental groups.
The role of static disorder in the eigenenergies of the system as
well as in the distribution of the angle $\phi$ will necessarily
yield an inhomogenously averaged signal from which the relevant information
must be carefully extracted. We anticipate this feature to add another
step of parameter fitting, but not change the results of our theory
dramatically. Furthermore, we have ignored the possibility of resolving
the vibronic structure accompanying each of the four resonances in
the 2D-ES. If this were to happen, it might be wiser to take the approach
of Cina and coworkers \cite{cinakilinhumble,cinabiggs1,cinabiggs2}
to consider the evolution of the nuclear wavepackets for a few modes
strongly coupled to the system, and maybe regard the rest of the modes
as a bosonic bath. This possibility would require an exponential increase
of experimental resources \cite{resources}, so either partial or
compressed sensing approaches \cite{rabitz,shabani} would be necessary.
Alternatively, by going back to the time-domain picture provided by
the authors in their previous work \cite{yuenzhou}, and applying
novel concepts of QPT for initially correlated states \cite{kavancesar,kavanpra,kavanarxiv},
a coarse grained and consistent tomographic protocol could be designed
to address this problem. Finally, it might be worth considering additional
nonlinear optical spectroscopic techniques such as considering the
analysis of both rephasing and non-rephasing signals \cite{cheng,readjpcb},
transient grating \cite{ppv}, pump probe \cite{khalil:362}, or phase
cycling of multipulse induced fluorescence \cite{tekavec:214307}
to investigate if they provide additional information for a more robust
QPT. 

Albeit this article not being exhaustive, we hope to have convinced
the reader that the QPT approach follows the spirit of MDOS in a very
natural way. By systematically studying excited state dynamics as
a quantum black box, an intriguing perspective on MDOS emerges that
allows the use of tools designed in the QIP community in order to
study excited state dynamics of condensed molecular systems. These
possibilities will be the subject of future studies.

\section{Acknowledgments}

We wish to acknowledge valuable discussions with Dr. Jacob Krich and
Dr. Paul Peng on nonlinear spectroscopy, Seungwan Ryu on time-bandwidth
issues, and Patrick Wen, Dylan Arias, Gabriela Schlau-Cohen, and Eleonora
de Re for insights on experimental realizations of QPT. This work
was funded by DARPA-QUBE as well as the Center for Excitonics.

\section*{APPENDIX: Derivation of Eqs. (\ref{eq:transformation}), (\ref{eq:hermiticity-1}),
and (\ref{eq:trace})}

\emph{\noun{Proof of Eq. (\ref{eq:transformation}).--}} Consider
a system $S$ interacting with a bath $B$. The total density matrix
of the composite object is $\rho_{total}$, whereas the reduced one
for the system and the bath are $\rho$ and $\rho_{B}$, respectively.
Suppose that the total initial state is a tensor product of the form: 

\begin{equation}
\rho_{total}(0)=\rho(0)\otimes\rho_{B}(0).\label{eq:tensor}\end{equation}

Where $\rho_{B}(0)$ is assumed to be fixed at:

\begin{equation}
\rho_{B}(0)=\sum_{\beta}p_{\beta}|e_{\beta}\rangle\langle e_{\beta}|,\label{eq:rho_bath}\end{equation}
 with $p_{\beta}\geq0$, for every initial state $\rho(0)$ of the
system.

At time $T$, the state of the composite object is simply a rotation
of the initial state (it is a closed system):

\begin{equation}
\rho_{total}(T)=U(T)\rho_{total}(0)U^{+}(T).\label{eq:unitary}\end{equation}
Here, $U(T)=\mathcal{T}(e^{-i\int_{0}^{T}H_{total}(t')dt'})$ is the
propagator for the entire object, where $\mathcal{T}$ is the time-ordering
operator, and $H_{total}$ is given by:

\begin{equation}
H_{total}=H_{S}+H_{B}+H_{SB},\label{eq:hamiltonian total}\end{equation}
where $H_{S},H_{B},H_{SB}$ are terms in the Hamiltonian that depend
only on $S$, on $B$, or on degrees of freedom of both, respectively.
Taking the trace of Eq. (\ref{eq:unitary}) with respect to the states
of $B$ yields $\rho(T)$:

\begin{equation}
\rho(T)=\sum_{\alpha\beta}E_{\alpha\beta}(T)\rho(0)E_{\alpha\beta}^{+}(T)\label{eq:OSR}\end{equation}
where:

\begin{equation}
E_{\alpha\beta}(T)=\sqrt{p_{\beta}}\langle e_{\alpha}|U(T)|e_{\beta}\rangle,\label{eq:Kraus}\end{equation}
is a Kraus operator and Eq. (\ref{eq:OSR}) is often known as the
\emph{operator sum representation} \cite{Choi1975285,sudarshan}.
By identifying:

\begin{equation}
\chi_{abcd}(T)=\sum_{\alpha\beta}[E_{\alpha\beta}(t)]_{ac}[E_{\alpha\beta}^{+}(t)]_{db},\label{eq:chi}\end{equation}
we have shown the equivalence between Eq. (\ref{eq:OSR}) and (\ref{eq:transformation}).

\begin{flushright}
$\Box$
\par\end{flushright}

~

\emph{\noun{Proof of Eq. (\ref{eq:trace}).--}}As explained in Section
II and in the previous proof, Eq. (\ref{eq:transformation}) holds
for any proper or improper state $\rho(0)$ in the Liouville space
of the system. With this in mind, Eq. (\ref{eq:trace}) is straightforward
to prove. We want to enforce that Eq. (\ref{eq:transformation}) preserves
trace throughout the evolution in time $T$. If $\rho(0)=|k\rangle\langle l|$,
then it must be that $Tr(\rho(T))=Tr(\rho(0))=\delta_{kl}$. Writing
the elements of the initial density matrix as $\rho_{cd}(0)=\delta_{ck}\delta_{dl}$,
we immediately obtain the condition we want to prove:

\begin{eqnarray}
Tr(\rho(T)) & = & \sum_{a}\delta_{ab}\sum_{cd}\chi_{abcd}\delta_{ck}\delta_{dl},\nonumber \\
\delta_{kl} & = & \chi_{aakl}.\label{eq:trace proof}\end{eqnarray}

\begin{flushright}
$\Box$
\par\end{flushright}

~

\emph{\noun{Proof of Eq. (\ref{eq:hermiticity-1}).-- }}Again, we
exploit the fact that Eq. (\ref{eq:transformation}) is true for any
$\rho(0)$.\emph{\noun{ }}

First, we consider a population as the initial state, i.e. $\rho(0)=|k\rangle\langle k|$
for any $k$. Note that $\rho(0)$ is a proper density matrix, and
hence Hermitian. Then, 

\begin{eqnarray}
\rho_{ab}(T) & = & \sum_{cd}\chi_{abcd}(T)\delta_{ck}\delta_{dk}\nonumber \\
 & = & \chi_{abkk}(T),\label{eq:hermiticity proof 1}\end{eqnarray}

\begin{eqnarray}
\rho_{ba}(T) & = & \sum_{dc}\chi_{bacd}(T)\delta_{ck}\delta_{dk}\nonumber \\
 & = & \chi_{bakk}(T).\label{eq:hermiticity proof 2}\end{eqnarray}
Since $\chi(T)$ maps proper density matrices to proper density matrices,
$\rho(T)$ must also be Hermitian, i.e. $\rho_{ba}(T)=\rho_{ab}^{*}(T)$.
It follows from Eqs. (\ref{eq:hermiticity proof 1}) and (\ref{eq:hermiticity proof 2})
that for any $a,b,k$:

\begin{equation}
\chi_{bakk}(T)=\chi_{abkk}^{*}(T).\label{eq:herm pop}\end{equation}
This proves Eq. (\ref{eq:hermiticity-1}) for the case $c=d$. Next,
we consider initial proper states of the form $\rho(0)=\frac{1}{2}(|k\rangle\langle k|+|k\rangle\langle l|+|l\rangle\langle k|+|l\rangle\langle l|)$.
By repeating the exercise above, we get:

\begin{equation}
\frac{1}{2}(\chi_{bakk}(T)+\chi_{bakl}(T)+\chi_{balk}(T)+\chi_{ball}(T))=\frac{1}{2}(\chi_{abkk}(T)+\chi_{abkl}(T)+\chi_{ablk}(T)+\chi_{abll}(T))^{*}\label{eq:sums herm}\end{equation}
Substituting the result Eq. (\ref{eq:herm pop}) into Eq. (\ref{eq:sums herm})
yields:

\begin{equation}
\chi_{bakl}(T)+\chi_{balk}(T)=\chi_{abkl}^{*}(T)+\chi_{ablk}^{*}(T)\label{eq:2sumsherm}\end{equation}
This result \emph{almost} proves Eq. (\ref{eq:hermiticity-1}) for
the case $c\neq d$. In order to complete the proof, we analyze another
class of initial proper states $\rho(0)=\frac{1}{2}(|k\rangle\langle k|+i|k\rangle\langle l|-i|l\rangle\langle k|+|l\rangle\langle l|)$.
The result from the same manipulations is that:

\begin{equation}
i\chi_{bakl}(T)-i\chi_{balk}(T)=-i\chi_{abkl}^{*}(T)+i\chi_{ablk}^{*}(T)\label{eq:2sumshermbis}\end{equation}
Comparing Eqs. (\ref{eq:sums herm}) and (\ref{eq:2sumsherm}) yields:

\begin{eqnarray}
\chi_{bakl}(T) & = & \chi_{ablk}^{*}(T)\nonumber \\
\chi_{balk}(T) & = & \chi_{abkl}^{*}(T)\label{eq:herm coh}\end{eqnarray}
for any $a,b,k,l$. Eqs. (\ref{eq:herm pop}) and (\ref{eq:herm coh})
together yield Eq. (\ref{eq:hermiticity-1}).

\bibliographystyle{unsrt}
\bibliography{quantumChem2}

\end{document}